\documentclass[twocolumn]{aastex631}

\usepackage{multirow}
\usepackage{appendix}

\newcommand{\kms}{\rm km~s^{-1}}

\newcommand{\Msun}{{\rm M}_{\odot}}
\newcommand{\dn}{{\rm D}_{n}4000}

\submitjournal{ApJ}

\begin{document}

\title{HectoMAP: The Complete Redshift Survey (Data Release 2)}

\correspondingauthor{Jubee Sohn}
\email{jbsohn@astro.snu.ac.kr}

\author[0000-0002-9254-144X]{Jubee Sohn}
\affil{Astronomy Program, Department of Physics and Astronomy, Seoul National University, 1 Gwanak-ro, Gwanak-gu, Seoul 08826, Republic of Korea} 
\affil{Smithsonian Astrophysical Observatory, 60 Garden Street, Cambridge, MA 02138, USA} 

\author[0000-0002-9146-4876]{Margaret J. Geller}
\affil{Smithsonian Astrophysical Observatory, 60 Garden Street, Cambridge, MA 02138, USA} 

\author[0000-0003-3428-7612]{Ho Seong Hwang} 
\affil{Astronomy Program, Department of Physics and Astronomy, Seoul National University, 1 Gwanak-ro, Gwanak-gu, Seoul 08826, Republic of Korea} 
\affil{SNU Astronomy Research Center, Seoul National University, 1 Gwanak-ro, Gwanak-gu, Seoul 08826, Republic of Korea}

\author[0000-0002-1311-4942]{Daniel G. Fabricant}
\affiliation{Smithsonian Astrophysical Observatory, 60 Garden Street, Cambridge, MA 02138, USA} 

\author[0000-0001-6161-8988]{Yousuke Utsumi} 
\affil{SLAC National Accelerator Laboratory, Menlo Park, CA 94025, USA; Kavli Institute for Particle Astrophysics and Cosmology, Stanford University, Stanford, CA 94305, USA} 

\author[0000-0003-4797-5246]{Ivana Damjanov} 
\affil{Department of Astronomy and Physics, Saint Mary's University, 923 Robie Street, Halifax, NS B3H 3C3, Canada}

\begin{abstract}
HectoMAP is a dense redshift survey of 95,403 galaxies based primarily on MMT spectroscopy with a median redshift $z = 0.345$. The survey covers 54.64 square degrees in a 1.5$^\circ$ wide strip across the northern sky centered at a declination of 43.25$^\circ$. We report the redshift, the spectral indicator $\dn$, and the stellar mass. The red selected survey is 81\% complete for 55,962 galaxies with $(g-r) > 1$ and $r <20.5$; it is 72\% complete for 32,908 galaxies with $(g-r) > 1$, $(r-i) > 0.5$ and $20.5 < r < 21.3$. Comparison of the survey basis SDSS photometry with the HSC-SSP photometry demonstrates that HectoMAP provides complete magnitude limited surveys based on either photometric system. We update the comparison between the HSC-SSP photometric redshifts with HectoMAP spectroscopic redshifts; the comparison demonstrates that the HSC-SSP photometric redshifts have improved between the second and third data releases. HectoMAP is a foundation for examining the quiescent galaxy population (63\% of the survey), clusters of galaxies, and the cosmic web. HectoMAP is completely covered by the HSC-SSP survey, thus enabling a variety of strong and weak lensing investigations. 
\end{abstract}

\section{INTRODUCTION} \label{sec:intro}

The universe of redshift surveys has undergone enormous expansion since the early surveys (e.g., \citealp{Davis82, Geller89, Shectman96}). Digital detectors enabled these pioneering surveys. Ever increasing multiplexing enabled by the combination of robotics and fiber optics underlies current and upcoming impressive projects: GAMA \citep{Robotham10, Liske15}, WEAVE \citep{Dalton12}, Subaru/PFS \citep{Takada14}, 4MOST \citep{Richard19}, DEVILS \citep{Davies18}, MOONRISE \citep{Maiolino20} and DESI \citep{Hahn22}.

Hectospec, a 300-fiber robotic instrument \citep{Fabricant05} mounted on the 6.5-meter MMT has contributed significantly to redshift surveys covering the range $0.2 < z < 0.8$ \citep{Kochanek12, Geller14, Geller16, Damjanov18}. SHELS \citep{Geller16} is a magnitude-limited redshift survey of two Deep Lensing Survey fields \citep{Wittman06} without any color selection. \citet{Damjanov22b} use SHELS to calibrate the much larger red-selected HectoMAP survey. HectoMAP is a dedicated redshift survey using Hectospec to study galaxies in the redshift range $0.2 < z < 0.6$. We follow the \citet{Damjanov22b} approach in further highlighting the completeness of the HectoMAP sample of quiescent galaxies.

The full HectoMAP survey covers 54.64 square degrees in a 1.5$^{\circ}$ wide strip centered at a declination of 43.25$^{\circ}$. The red-selected survey is designed to provide dense sampling of the quiescent galaxy population \citep{Damjanov22a}, clusters of galaxies \citep{Sohn18b, Sohn18a, Sohn21b}, and the cosmic web \citep{Hwang16}. The average survey density is $\sim 1800$ galaxies deg$^{-2}$ and the median redshift $z = 0.345$. \citet{Pizzardo22} demonstrate that this sampling positions HectoMAP as the foundation for a direct measurement of the galaxy cluster growth rate over the last 7 gigayears of cosmic history. 

\citet{Sohn21a} is the first HectoMAP data release (DR1 hereafter). DR1 covers 8.7 deg$^{2}$ and includes 17,313 galaxies. As demonstrations of the broad scientific applications of HectoMAP, the paper includes an assay of redMaPPer clusters \citep{Rykoff16} in the region and a test of the HSC DeMP photometric redshifts \citep{Hsieh14}.

The availability of HSC-SSP $grizy$ photometry \citep{Aihara22} covering the entire HectoMAP region also distinguishes the survey and enables investigations that combine two of the powerful tools of modern cosmology, redshift surveys and weak lensing. HectoMAP promises a platform for these combined studies ranging from the masses of galaxies to the relationship between the galaxy and dark matter distributions on large scales. \citet{Damjanov22a} also highlight the power of the completeness of the HectoMAP quiescent sample combined with sizes from HSC-SSP photometry for elucidating the evolution of the size-mass relation for the quiescent population in the redshift range $0.2 < z < 0.6$.

Here we describe the full HectoMAP redshift survey data including 95,403 galaxies in the entire 54.64 square degree region of HectoMAP; 88294 ($\sim 93\%)$ of these redshifts are MMT Hectospec observations. Selection for the complete HectoMAP sample is based on SDSS DR16 \citep{SDSSDR16} photometry: $(g - r) > 1$ for $r \leq 20.5$. For $20.5 < r \leq 21.3$, we include an additional constraint, $(r - i) > 0.5$. We also provide $\dn$ and the stellar mass for most of galaxies with spectroscopy. 

We describe the data in Section \ref{sec:data}. Section \ref{sec:phot} reviews the photometric basis for the survey. Section \ref{sec:spec} describes the HectoMAP spectroscopy, including the MMT spectroscopy (Section \ref{sec:MMTspec}) that provides the vast majority of the spectroscopic data. Section \ref{sec:result} includes the survey catalog. Sections \ref{sec:dn} and \ref{sec:mass} describe the $\dn$ and stellar mass, respectively. We include the full HectoMAP catalog in Section \ref{sec:cat}. The red-selected survey and the subset of quiescent galaxies are highly complete (Section \ref{sec:complete}). Section \ref{sec:HSC} compares the SDSS and HSC-SSP photometry in the region, and Section \ref{sec:zphot} compares updated photometric redshifts based on HSC-SSP DR3 photometry \citep{Aihara22} with HectoMAP spectroscopic redshifts. Section \ref{sec:apps} highlights past and future applications of HectoMAP including a brief discussion of strong lensing clusters in the region. We conclude in Section \ref{sec:conclusion}. We use the Planck cosmological parameters: $H_{0} = 67.74~{\rm km~s^{-1}~Mpc^{-1}}$, $\Omega_{m} = 0.3089$, and $\Omega_{\Lambda} = 0.6911$ throughout.

\section{THE DATA} \label{sec:data}

HectoMAP is a large, deep, dense redshift survey enabling investigation of the quiescent galaxy population, clusters of galaxies, and large-scale structure in the redshift range $0.2 < z < 0.6$. HectoMAP covers a 54.64 square degree region within a narrow strip across the northern sky: $200 < $ R.A. (deg) $< 250$ and $42.5 <$ Decl. (deg) $< 44.0$. Data Release 1 (DR1) \citep{Sohn21a} includes photometric and spectroscopic properties of galaxies in the region $242 < $ R.A. (deg) $< 250$. Here, we include the complete redshift sample for HectoMAP.

We first describe the photometric and spectroscopic basis for the HectoMAP spectroscopic catalog. The photometric data (Section \ref{sec:phot}) include the survey basis from SDSS (Section \ref{sec:SDSSphot}) and the later much deeper Subaru HSC/SSP photometry (Section \ref{sec:HSCphot}). We next describe the HectoMAP spectroscopy (Section \ref{sec:spec}). SDSS spectroscopy provides a low redshift sample (Section \ref{sec:SDSSspec}). MMT observations with Hectospec provide the vast majority of the spectroscopy (Section \ref{sec:MMTspec}). 

\subsection{Photometry} \label{sec:phot}
\subsubsection{SDSS Photometry} \label{sec:SDSSphot}

Galaxy selection for the HectoMAP spectroscopic survey is based on SDSS DR16 photometry \citep{SDSSDR16}. We selected galaxies with $probPSF = 0$, where $probPSF$ gives the probability that the object is a star\footnote{https://classic.sdss.org/dr3/algorithms/classify.html$\sharp$photo$\_$class}. We applied additional object selection criteria based on the SDSS photometric flags to remove spurious sources, including stellar bleeding, suspicious detections, objects with deblending problems, and objects without proper Petrosian photometry. There are 212,120 galaxies with $r \leq 21.3$, the magnitude limit of HectoMAP. 

We base the survey on SDSS $ugriz$ Petrosian magnitudes after foreground extinction correction. We also obtain colors based on model magnitudes following our previous spectroscopic surveys (e.g., F2, \citealp{Geller14}). Petrosian magnitudes are galaxy fluxes measured within a circular aperture with a radius defined by the azimuthally averaged light profile \citep{Petrosian76}. SDSS photometry provides galaxy fluxes resulting from fits of de Vaucouleur and exponential models. Model magnitudes select the better of these two fits in the $r-$band as a basis for calculating the flux in all bands. Hereafter, $r_{petro,0}$ refers to the SDSS Petrosian magnitude corrected for foreground extinction. The SDSS model colors, $(g - r)_{model,0}$ and $(r - i)_{model,0}$, are also corrected for foreground extinction. Hereafter, we designate $r = r_{petro, 0}$, $(g - r) = (g - r)_{model,0}$, and $(r - i) = (r - i)_{model,0}$ in the text; we retain the full notation in the figure captions for clarity. 

Additionally, we compile the composite model (cModel) magnitudes. The cModel magnitudes are based on a linear combination of the best fitting exponential and de Vaucouleurs models \citep{Strauss02}. For galaxies, cModel magnitudes agree well with Petrosian magnitudes. We use cModel magnitudes for comparison with the much deeper HSC-SSP photometry (Section \ref{sec:HSC}) where similarly defined cModel magnitudes are available. 

\subsubsection{Subaru/HSC Photometry}\label{sec:HSCphot}

HectoMAP is covered by the HSC-SSP wide survey \citep{Miyazaki12}, which provides deep photometry to a $5\sigma$ depth of 26.5, 26.5, 26.2, 25.2, and 24.4 mag in the $g$, $r$, $i$, $z$, and $y$-bands for point sources, respectively \citep{Aihara22}. We use the HSC-SSP Public Data Release 3 (hereafter HSC-SSP DR3, \citealp{Aihara22}) to explore the relationship between the SDSS and Subaru photometry (Section \ref{sec:HSC}). The HSC-SSP DR3 includes the complete HectoMAP survey region in the $g, r$ and $z$ bands; the survey for $i$ and $y$ bands are $\sim 80\%$ complete. 

We use the `forced' DR3 catalog that lists forced photometry on coadded images. The forced DR3 photometry is based on the set of images generated with a local sky subtraction scheme \citep{Aihara22}. In the catalog, there are two types of photometry: Kron and cModel magnitudes. We use cModel magnitudes \citep{Bosch18} for direct comparison with the cModel magnitudes from SDSS. The cModel photometry yields unbiased color estimates \citep{Huang18a}. 

We match the HSC-SSP catalog to the SDSS photometric catalog with a matching tolerance of $1\arcsec$. Most of the SDSS galaxies have HSC counterparts (93\%). The SDSS objects missing from HSC photometry are mostly near bright saturated stars in the HSC images. We discuss the comparison between HSC and SDSS photometry in Section \ref{sec:HSC}. 

\subsection{Spectroscopy} \label{sec:spec}

The HectoMAP survey includes redshifts for 95,403 unique galaxies; 88,294 of these redshifts are MMT/Hectospec observations (Section \ref{sec:MMTspec}). Particularly at redshifts $z < 0.2$, SDSS complements the HectoMAP data (Section \ref{sec:SDSSspec}). The NASA Extragalactic Database adds only a few (375) unique redshifts in the HectoMAP survey region. 

Table \ref{tab:summary} summarizes spectroscopic redshifts from different sources in various subsamples. The first line of Table \ref{tab:summary} includes the stars discussed in Appendix. All other entries are samples of galaxies.
 
\begin{deluxetable*}{lcccccc}
\label{tab:summary}
\tabletypesize{\tiny}
\tablecaption{HectoMAP Survey Subsets}
\tablecolumns{7}
\tablewidth{0pt}
\tablehead{\colhead{Subsample} & \colhead{$N_{phot}$} & 
\colhead{$N_{spec, MMT}$} & \colhead{$N_{spec, SDSS/BOSS}$} & \colhead{$N_{spec, NED}$} & 
\colhead{$N_{spec, total}$} & \colhead{$f_{complete}$}}
\startdata
Entire sample with spectroscopy$^{*}$ ($r \leq 23$) & \nodata &  94114  & 19538  &  464 & 114116 & \nodata \\
Galaxies with $r \leq 23$                           & \nodata &  88294  &  6734  &  375 &  95403 & \nodata \\
Galaxies within $21.3 < r \leq 23$                  & \nodata &   6095  &  4555  &    0 &  10650 & \nodata \\
Main ($r \leq 21.3$))                               &  212120 &  82610  &  5560  &  326 &  88496 &   41.72 \\
Main ($(g-r) > 1.0$)                                &  108555 &  69788  &  2777  &   54 &  72619 &   66.90 \\
Main ($(g-r) > 1.0$ \& $(r-i) > 0.5$)               &   65729 &  52985  &  1923  &    5 &  54913 &   83.54 \\
Main ($(g-r) \leq 1.0$)                             &  103565 &  12822  &  2783  &  205 &  15810 &   15.27 \\
Bright ($r \leq 20.5$)                              &  107166 &  54983  &  4614  &  312 &  59909 &   55.90 \\
Bright ($r \leq 20.5$)                              &   55963 &  43435  &  1921  &   51 &  45407 &   81.14 \\
Bright ($(g-r) > 1.0$ \& $(r-i) > 0.5$)             &  32821  &  30072  &  1069  &    3 &  31144 &   94.89 \\
Bright ($(g-r) \leq 1.0$)                           &   51203 &  11548  &  2693  &  195 &  14436 &   28.19 \\
Faint ($20.5 < r \leq 21.3$)                        &  104954 &  27627  &   946  &   14 &  28587 &   27.24 \\
Faint ($(g-r) > 1.0$)                               &   52592 &  26353  &   856  &    3 &  27212 &   51.74 \\
Faint ($(g-r) > 1.0$ \& $(r-i) > 0.5$)              &   32908 &  22913  &   854  &    2 &  23769 &   72.23 \\
Faint ($(g-r) \leq 1.0$)                            &   52362 &   1274  &    90  &   11 &   1375 &    2.63 
\enddata 
\tablenotetext{*}{This sample includes stars and galaxies with spectroscopy.}
\end{deluxetable*}

\subsubsection{SDSS Spectroscopy}\label{sec:SDSSspec}

The SDSS main galaxy sample is a complete spectroscopic survey ($\sim 90 - 95\%$ complete, \citealp{Strauss02, Lazo18}) for galaxies with $r < 17.77$. We include 3932 unique SDSS main galaxy sample redshifts in the HectoMAP catalog. The SDSS spectra cover the wavelength range 3700 - 9100 {\rm \AA} with a typical spectral resolution of $\sim 3~{\rm \AA}$. The typical uncertainty of the SDSS redshifts we use is $8~\kms$.

There are also 2802 unique SDSS/BOSS redshifts in the HectoMAP region. BOSS spectra cover the wavelength range $3600 - 10400 {\rm \AA}$. Because BOSS typically targets fainter and higher redshift objects, the uncertainty in the BOSS redshifts is generally larger ($\sim 39~\kms$) than for the SDSS redshifts. 

\begin{figure*}
\centering
\includegraphics[scale=0.37]{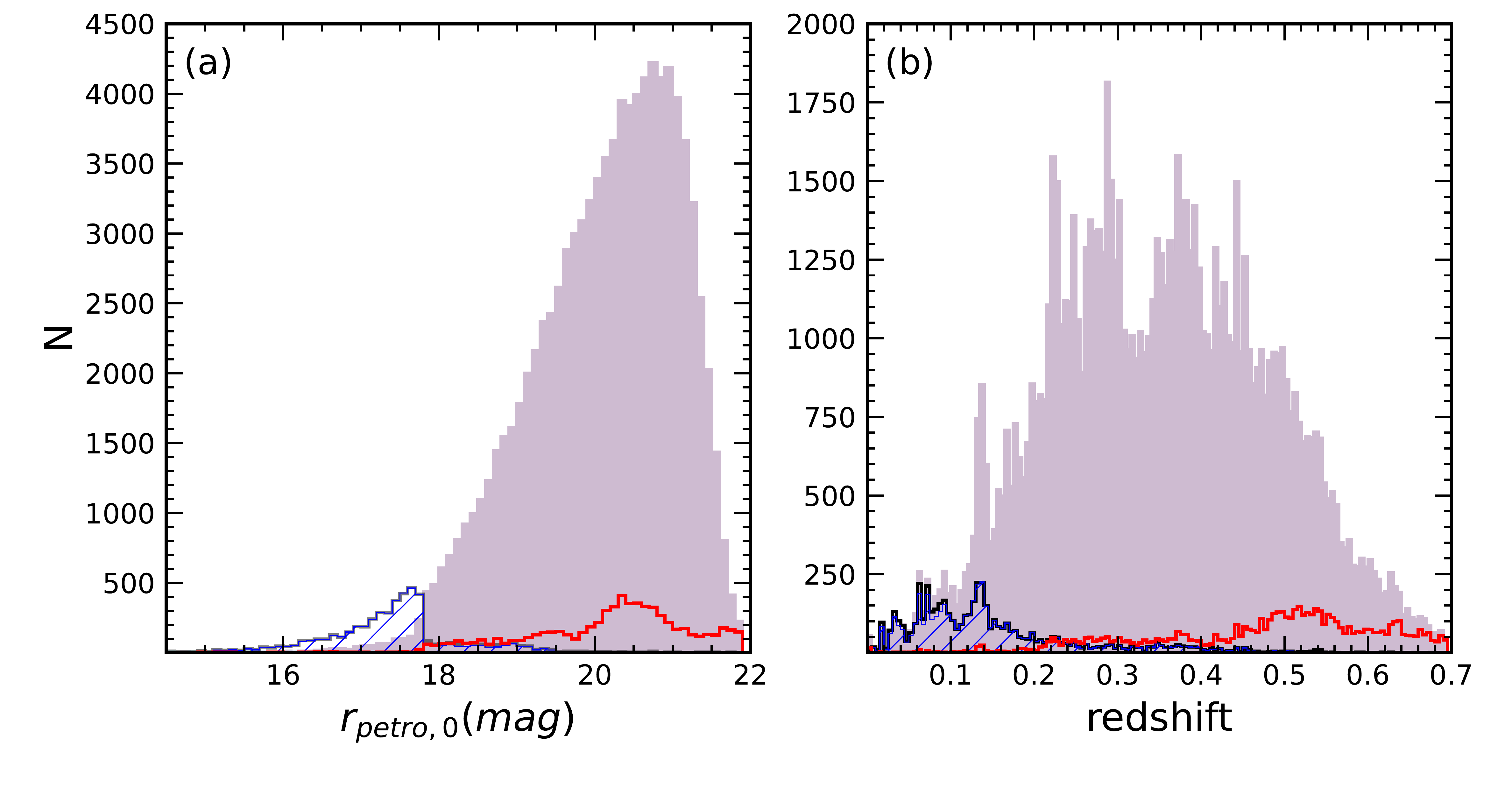}
\caption{(a) Distribution of $r_{petro, 0}$ for MMT/Hectospec (purple filled), SDSS (blue hatched), BOSS (red open) galaxies in HectoMAP. (b) Same as (a), but for the redshift distributions. }
\label{fig:magz}
\end{figure*}

The purple histograms in Figures \ref{fig:magz} (a) and (b) show the distributions of $r-$band magnitudes and redshifts for the MMT/Hectospec galaxies (Section \ref{sec:MMTspec}) in HectoMAP. Blue-hatched and red-open histograms display the small contribution to the overall survey from unique galaxies in SDSS and BOSS, respectively. The SDSS Main Galaxy Sample is limited to $z < 0.2$. Less than 1\% of the entire HectMAP spectroscopic sample is from SDSS. BOSS reaches fainter magnitudes ($r > 18$) and thus covers a wider redshift range; BOSS contributes $\sim 3\%$ of the HectoMAP spectroscopy. 

\subsubsection{MMT/Hectospec Spectroscopy}\label{sec:MMTspec}

We used MMT/Hectospec \citep{Fabricant05} to obtain spectroscopy for most of the galaxies in HectoMAP. Hectospec has 300 fibers each with a $1.5\arcsec$ aperture. A single observation generally acquires spectra for $\sim 250$ galaxies over the 1 degree diameter field of view of the instrument; the remaining fibers observe the sky. The 270 mm$^{-1}$ Hectospec grating yields a typical resolution of $\sim 6~{\rm \AA}$ and covers the wavelength range 3700 - 9100 {\rm \AA}. We made HectoMAP observations from 2009 January to 2019 May. 

HectoMAP is a red-selected survey. We targeted brighter galaxies with $r \leq 20.5$ and $(g-r) > 1.0$. For fainter galaxies with $20.5 < r \leq 21.3$, we applied an additional color selection $(r-i) > 0.5$ to complete the survey within the limited available telescope time. Conducting a magnitude-limited survey without color selection was prohibitive. We discuss the survey completeness in Section \ref{sec:complete}. 

We reduced the Hectospec spectra using the standard pipeline, HSRED v2.0. We derive redshifts based on the cross-correlation tool, RVSAO \citep{Kurtz98}. RVSAO yields the cross-correlation score ($R_{XC}$). Following previous Hectospec surveys, we take the $R_{XC} > 3$ (see Figure 3 of \citet {Sohn21a}) as a reliable redshift \citep{Rines16, Sohn21a}. The typical redshift uncertainty in a Hectospec redshift is $\sim 39~\kms$, comparable with the uncertainty in BOSS redshifts.

There are 924/4496 galaxies that have both Hectospec and SDSS/BOSS spectra. The Hectospec redshifts are slightly {\it lower} than the SDSS/BOSS redshifts: $c \Delta (z_{SDSS/BOSS} - z_{Hecto}) / (1 + z_{Hecto}) = 42~\kms$. \citet{Sohn21a} explored this issue in detail by showing the redshift difference as a function of the signal-to-noise (S/N) ratio of the spectra (see their Figure 6). Here, we note that the redshift difference is comparable to the typical uncertainty in a Hectospec redshift. The systematic difference does not affect any of the analysis we carry out here.

\section{The HectoMAP Spectroscopic Catalog} \label{sec:result}

Based on the photometric and spectroscopic data for the HectoMAP region, we derive several spectroscopic properties of the galaxies. We outline the spectroscopic properties including $\dn$ (Section \ref{sec:dn}), stellar mass ($M_{*}$, Section \ref{sec:mass}), and K-correction (Section \ref{sec:kcor}). We describe the complete HectoMAP spectroscopic DR2 catalog in Section \ref{sec:cat}.

\subsection{$\dn$} \label{sec:dn}

We derive the spectral indicator $\dn$ that measures the flux ratio around the $4000~{\rm \AA}$ break. The $\dn$ index is a useful and robust tracer of the age of stellar population of galaxies. For example, \citet{Kauffmann03} show that $\dn$ increases monotonically after a burst of star formation (see also \citealp{Zahid19}). This index is also useful for identifying the quiescent galaxy population (e.g., \citealp{Vergani08, Zahid16, Sohn17a, Sohn17b, Damjanov22a, Hamadouche22}). We use $\dn$ to select quiescent galaxies in HectoMAP. 

We compute the flux ratio between two spectral windows $3850 - 3950~{\rm \AA}$ and $4000 - 4100~{\rm \AA}$ following the definition from \citet{Balogh99}. The median signal-to-noise ratio at $3850 - 4100 {\rm \AA}$ is $\sim 4.5$. We measure the $\dn$ for 99\% of HectoMAP objects with spectroscopy. The missing $\dn$ objects are the few galaxies with a NED redshift where we do not have a spectrum. Table \ref{tab:dn} summarizes the fraction of galaxies with $\dn$ measurements and with $\dn > 1.5$. 

\begin{figure*}
\centering
\includegraphics[scale=0.35]{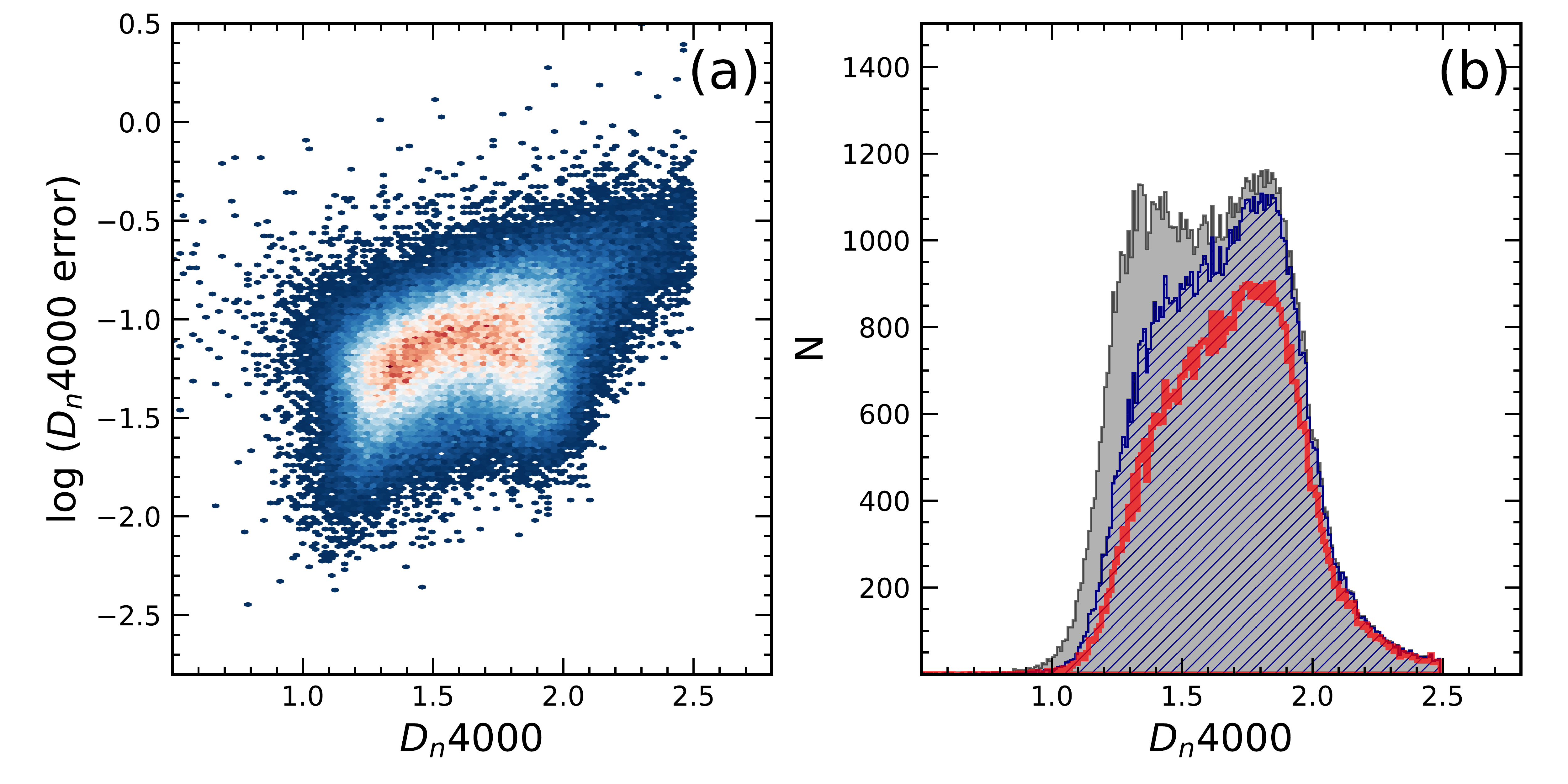} 
\caption{(a) $\dn$ absolute error vs. $\dn$ for HectoMAP galaxies. Redder color indicates higher number density. (b) The $\dn$ distribution of all HectoMAP galaxies (black filled histogram), galaxies with $(g-r) > 1.0$ (blue hatched), and galaxies with $(g-r) > 1.0$ and $(r-i) > 0.5$ (red open). }
\label{fig:dn}
\end{figure*}

Figure \ref{fig:dn} (a) shows the $\dn$ absolute error as a function of $\dn$. The typical $\dn$ uncertainties measured from SDSS, BOSS, and Hectospec spectra are 0.04, 0.10, and 0.07, respectively. We also compare the difference between $\dn$ measured from Hectospec and $\dn$ measured from SDSS/BOSS. The mean $\dn$ difference is very small, i.e., $\Delta \dn~{\rm(Hecto - SDSS/BOSS)} = -0.014 \pm 0.003$. 

Figure \ref{fig:dn} (b) displays the $\dn$ distribution. We overlay the red-selected galaxies with $(g-r) > 1.0$ (blue hatched histogram) and $(g-r) > 1.0$ and $(r-i) > 0.5$ (red open histogram), respectively. The $\dn$ distribution of HectoMAP galaxies is bimodal; the larger $\dn$ population is quiescent. We identify quiescent galaxies with $\dn \geq 1.5$ following previous MMT surveys. Overall, 63\% of HectoMAP galaxies are quiescent. This large quiescent fraction results from the red-selection of the survey. 

\begin{deluxetable*}{llcccc}
\label{tab:dn}
\tabletypesize{\tiny}
\tablecaption{Completeness of $\dn$ and Stellar Mass Measurements}
\tablecolumns{6}
\tablewidth{0pt}
\tablehead{\colhead{Subsample} & &
\colhead{$N_{sample}$} & 
\colhead{$f_{\dn}$} & 
\colhead{$f_{M_{*}}$} & 
\colhead{$f_{\dn > 1.5}$}}
\startdata
All Galaxies,         &                               &  95403 & 0.9996 & 0.9946 & 0.6314 \\
$r \leq 20.5$,        & $(g-r) > 1.0$                 &  45407 & 0.9994 & 0.9977 & 0.7699 \\
$r \leq 20.5$,        & $(g-r) > 1.0~ \& (r-i) > 0.5$ &  31147 & 0.9995 & 0.9986 & 0.8226 \\
$20.5 < r \leq 21.3$, & $(g-r) > 1.0$                 &  27212 & 0.9999 & 0.9988 & 0.6425 \\
$20.5 < r \leq 21.3$, & $(g-r) > 1.0~ \& (r-i) > 0.5$ &  23769 & 1.0000 & 0.9989 & 0.6872
\enddata 
\end{deluxetable*}

\subsection{Stellar Mass}\label{sec:mass}

We compute the stellar masses of galaxies as we did for previous MMT/Hectospec redshift surveys (e.g., \citealp{Geller14, Zahid16, Sohn17a, Sohn21a, Damjanov22a}). We use foreground-extinction corrected SDSS $ugriz$ model magnitudes to compute the stellar mass. We use the Le PHARE fitting code \citep{Arnouts99, Ilbert06} to fit the observed and model spectral energy distributions (SEDs). To generate model SEDs, we employ the stellar population synthesis models of \citet{BC03}. We assume a universal Chabrier initial mass function \citep{Chabrier03} and two metallicities. We also consider a set of models with exponentially declining star formation rates with e-folding times for the star formation ranging from 0.1 to 30 Gyr. We also consider the internal extinction range $E(B-V) = 0 - 0.6$ using the extinction law from \citet{Calzetti00}. The model SEDs are normalized to solar luminosity, and the ratio between the observed and model SEDs is the stellar mass. The median of the best-fit stellar mass distribution is the stellar mass we use. 

\begin{figure*}
\centering
\includegraphics[scale=0.22]{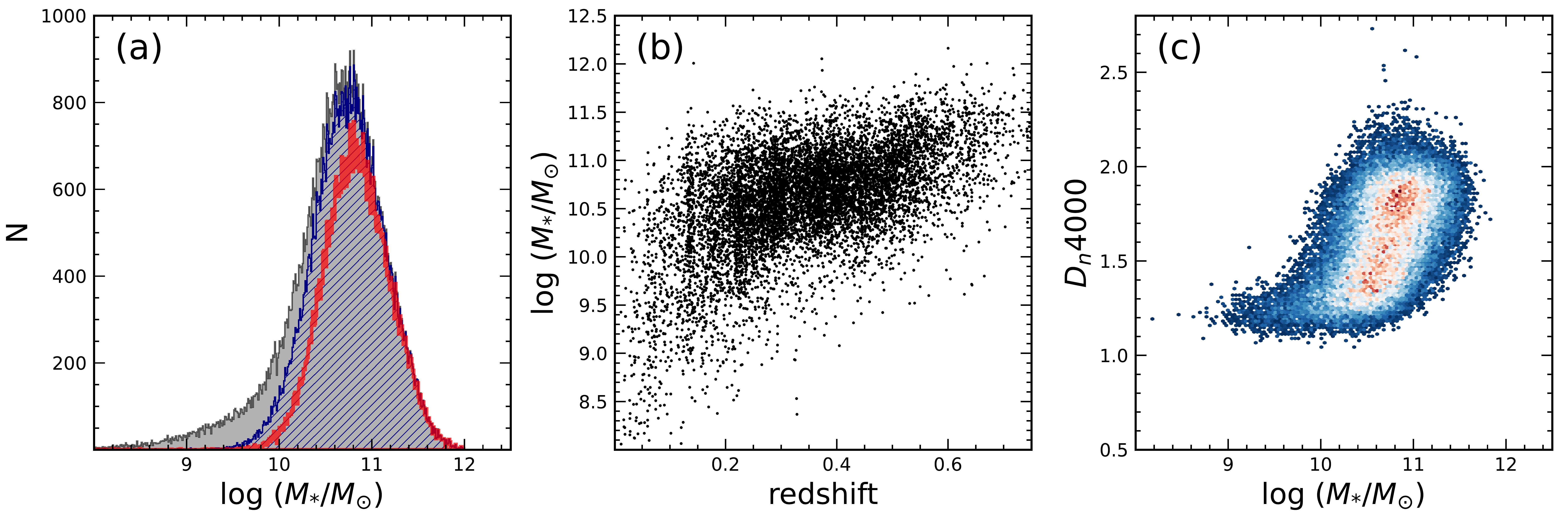}
\caption{(a) Stellar mass ($M_{*}$) distribution for all HectoMAP galaxies (black filled histogram), galaxies with $(g-r) > 1.0$ (blue hatched), and galaxies with $(g-r) > 1.0$ and $(r-i) > 0.5$ (red open). (b) $M_{*}$ as a function of redshift. (b) $\dn$ as a function of $M_{*}$. We show only 10\% of the HectoMAP objects for clarity in panels (b) and (c). }
\label{fig:mass}
\end{figure*}

Figure \ref{fig:mass} (a) shows the distribution of stellar masses for galaxies in the HectoMAP survey. The red selection of HectoMAP shifts the survey toward galaxies with generally high stellar mass. The stellar mass range is $8.0 < \log (M_{*} / M_{\odot}) < 12.0$. The typical stellar mass uncertainty is $\Delta \log (M_{*} / M_{\odot}) \sim 0.12 \pm 0.33$. We note that stellar mass estimates based on broad band photometry can also have systematic uncertainties of $\sim 0.3$ dex (e.g., \citealp{Conroy09}). 

Figure \ref{fig:mass} (b) displays the stellar mass as a function of redshift. As expected in a magnitude-limited survey, there are more low mass galaxies at low redshift. Figure \ref{fig:mass} (c) shows $\dn$ versus stellar mass. The low mass objects are mostly star-forming galaxies with low $\dn$. High mass objects generally have larger $\dn$. This behavior is well-known (e.g., \citealp{Kauffmann04, Blanton09}).

\subsection{K-correction}\label{sec:kcor}

\begin{figure*}
\centering
\includegraphics[scale=0.37]{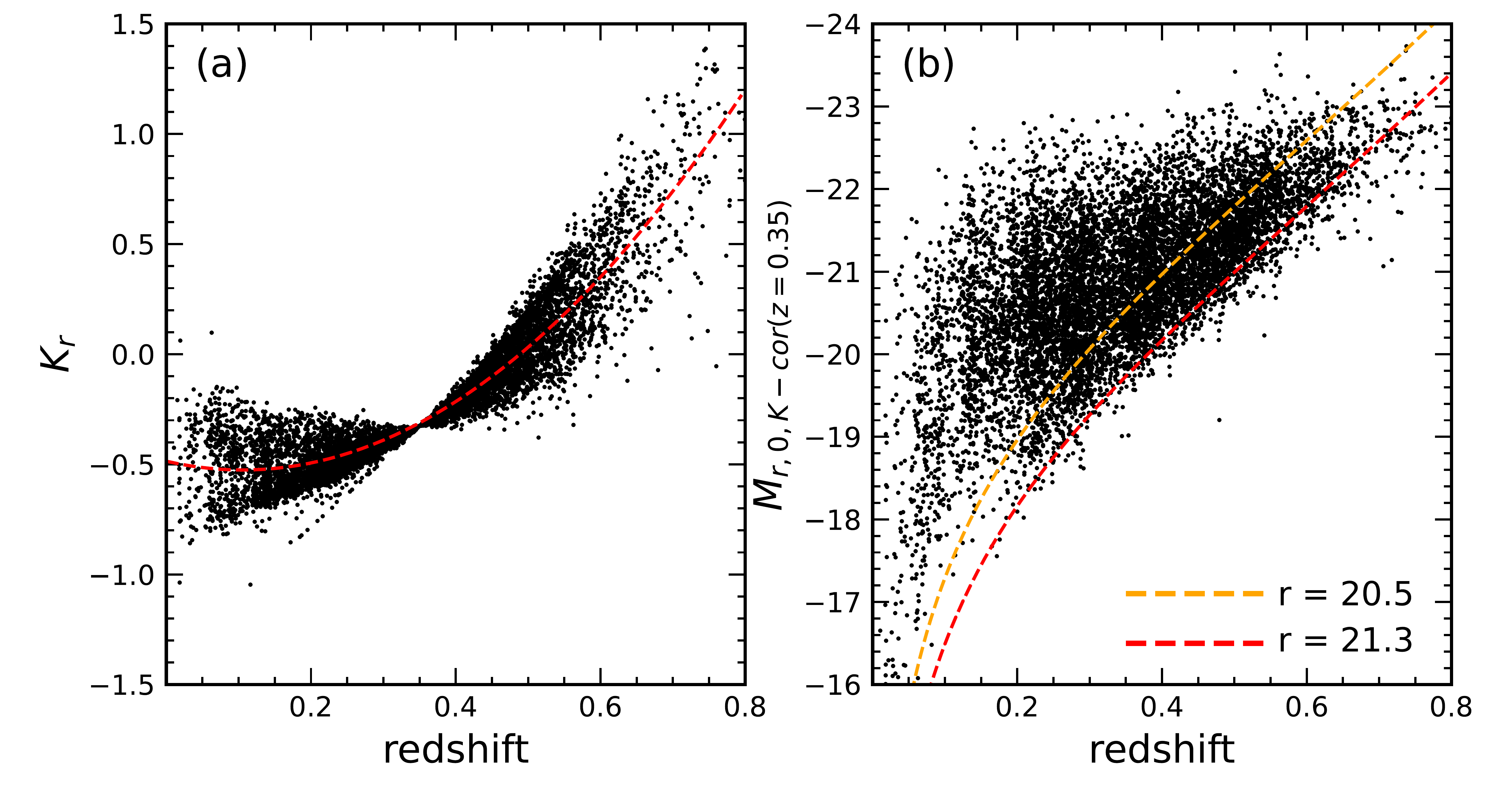}
\caption{(a) $r-$band K-correction as a function of redshift. Red dashed line shows the best-fit polynomial fit for the median K-correction. (b) The K-corrected $r-$band absolute magnitude as a function of redshift. Yellow and red dashed lines show the $r = 20.5$ and $r = 21.3$ magnitude limits shifted by the median K-correction, respectively. We plot only $10\%$ of randomly sampled HectoMAP galaxies for clarity. We  apply no color selection here.} 
\label{fig:kcor}
\end{figure*}

We calculate the K-correction based on the {\it kcorrect} IDL package \citep{Blanton07}. We derive the K-correction at $z = 0.35$, the median redshift of HectoMAP to minimize the K-correction we apply. Figure \ref{fig:kcor} (a) displays the K-correction distribution as a function of redshift. The K-correction at $z = 0.35$ is $-0.31$. The dashed line shows the median K-correction as a function of redshift. The best-fit median K-correction is:
\begin{equation}
K_{r} = -0.31 + 1.75 (z - 0.35) + 3.58 (z - 0.35)^{2}.
\end{equation}

Figure \ref{fig:kcor} (b) shows the K-corrected $r-$band absolute magnitude of HectoMAP galaxies as a function of redshift. Two dashed lines show the $r = 20.5$ and $r = 21.3$ magnitude limits shifted by the median K-correction (equation 1). These lines are useful for deriving volume-limited subsamples within HectoMAP. 

\subsection{The HectoMAP Spectroscopic Catalog} \label{sec:cat}

Table \ref{tab:cat} lists the 95,403 galaxies with spectroscopy in the full 54.64 deg$^{2}$ HectoMAP region. Table \ref{tab:star} in the Appendix lists the 6544 stars observed in the HectoMAP spectroscopic campaign. Among these stars, 2291 objects were identified as galaxies ($probPSF = 0$) in the SDSS photometric database.

Table \ref{tab:cat} includes the SDSS Object ID (Column 1), Right Ascension (Column 2), Declination (Column 3), the redshift and its uncertainty (Column 4), the SDSS $r-$band Petrosian magnitude and its uncertainty (Column 5), the K-corrected $r-$band absolute cModel magnitude (Column 6), the SDSS $r-$band cModel magnitude and its uncertainty (Column 7). We also list the HSC-SSP $r-$band cModel magnitude (Column 8). We use the HSC photometry as a basis for the Subaru/SDSS photometric comparison and for estimating the survey completeness based on HSC photometry. We list $\dn$ and its uncertainty (Column 9), the stellar mass and its uncertainty (Column 10), and the source of the spectroscopy (Column 11).  

\begin{figure*}[h!]
\centering
\includegraphics[scale=0.3]{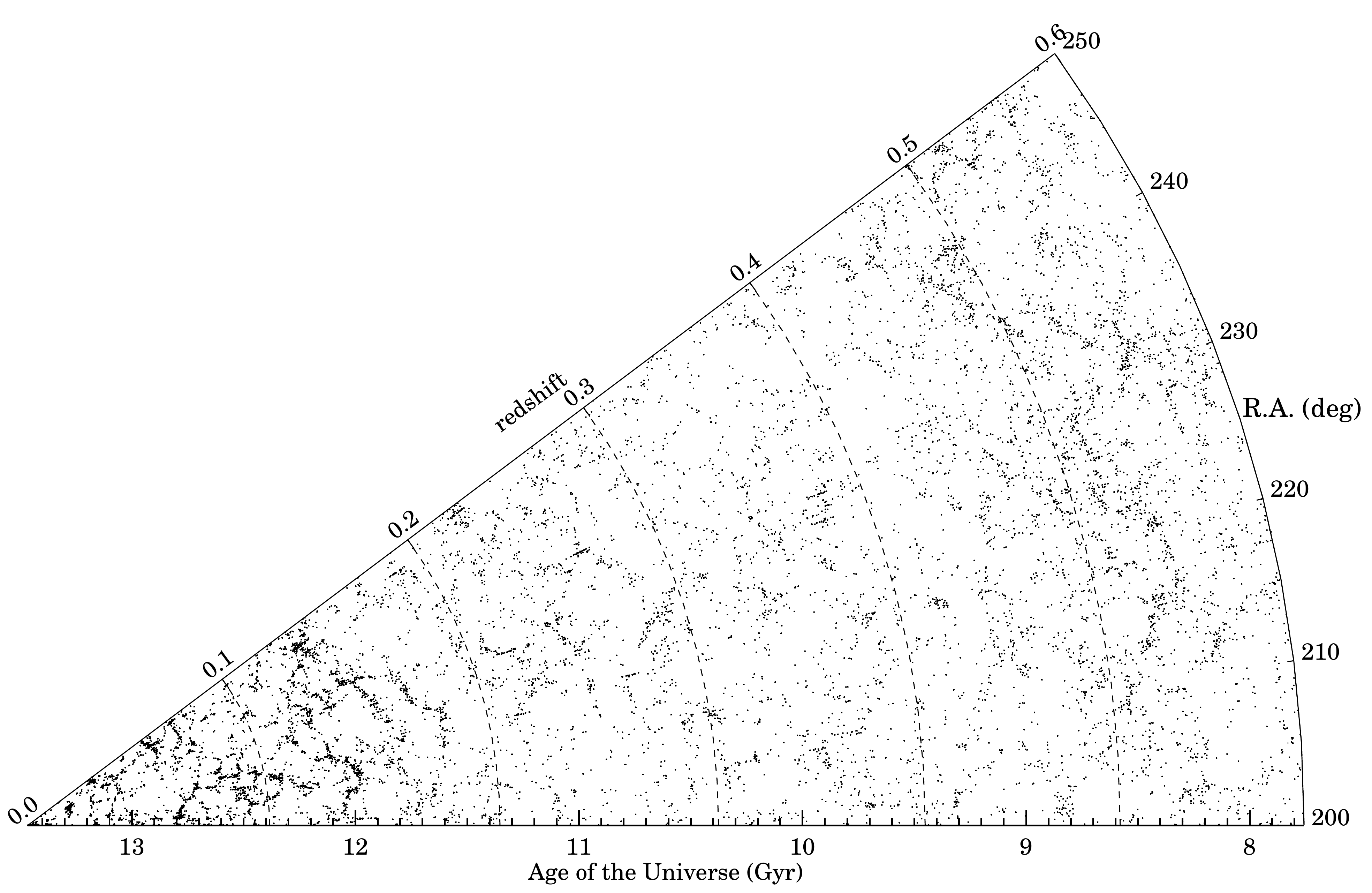}
\caption{HectoMAP cone diagram based only on SDSS/BOSS spectroscopic redshifts. Individual points show galaxies with SDSS/BOSS spectra. }
\label{fig:cone_sdss}
\end{figure*}

Figure \ref{fig:cone_sdss} displays a cone diagram for the HectoMAP region based only on SDSS/BOSS spectroscopy. The SDSS spectroscopic survey reveals the characteristic large-scale structure at $z < 0.18$; the survey density drops dramatically at higher redshift. The BOSS spectroscopic survey adds redshifts at $z > 0.5$. However, the number density is low. 

\begin{figure*}[h!]
\centering
\includegraphics[scale=0.3]{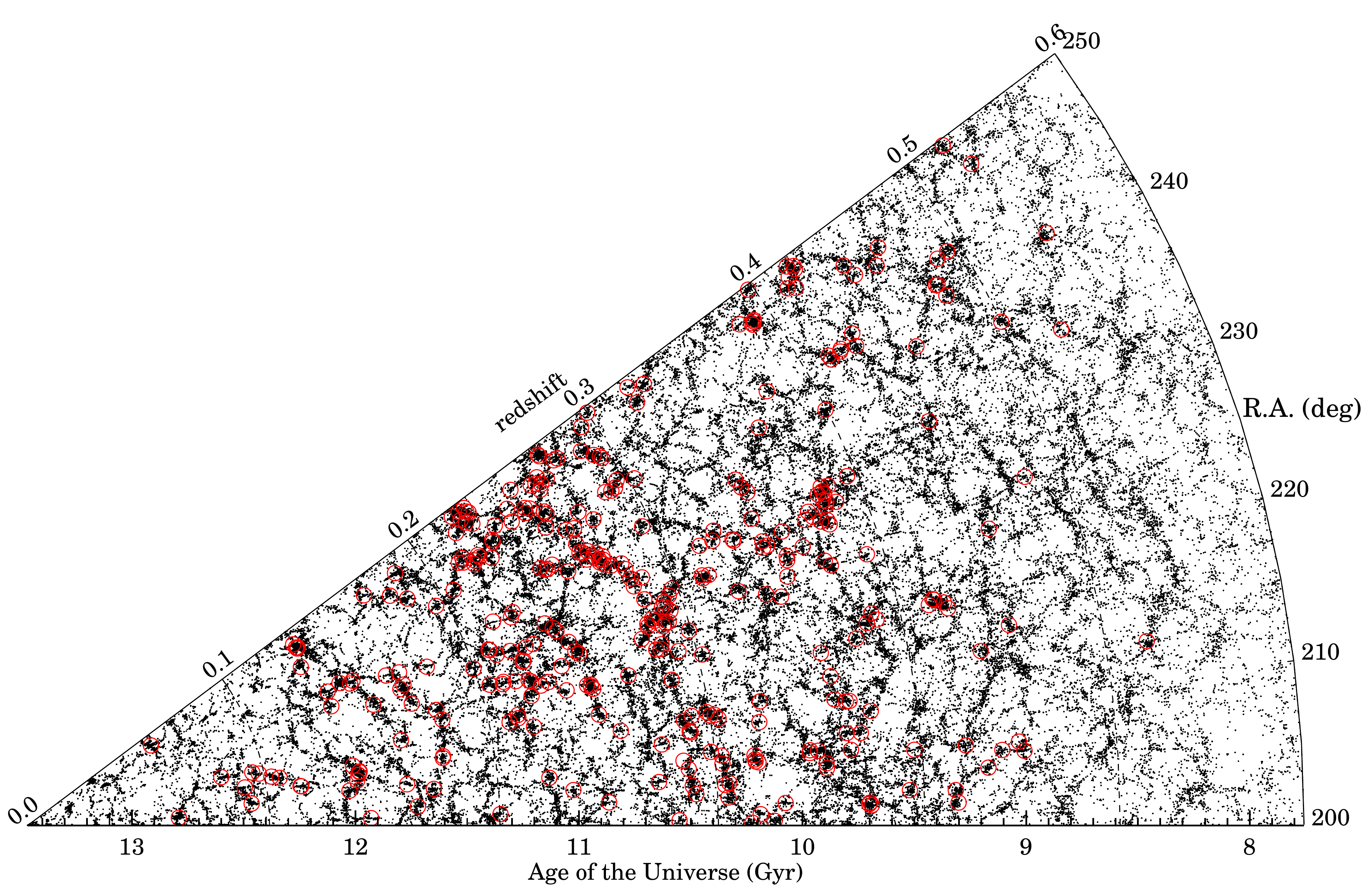} 
\caption{HectoMAP cone diagram for the entire HectoMAP survey. The characteristic filaments and voids of the cosmic web are obvious. Red points mark the centers of friends-of-friends galaxy clusters in \citet{Sohn21b}. }
\label{fig:cone}
\end{figure*}

Figure \ref{fig:cone} shows a cone diagram for all 95,403 galaxies in HectoMAP. In contrast with Figure \ref{fig:cone_sdss}, the characteristic filaments, walls and voids are strikingly evident. Expanding the figure also reveals the fingers that indicate the central regions of massive clusters. The red points mark the centers of the friends-of-friends clusters identified by \citet{Sohn21b}. The median redshift of the survey is $z = 0.345$. Beyond this redshift, the survey density declines as expected for a magnitude-limited survey. 

\begin{deluxetable*}{cccccccccccc}
\label{tab:cat}
\tabletypesize{\tiny}
\tablecaption{The HectoMAP Redshift Survey}
\tablecolumns{11}
\tablewidth{0pt}
\tablehead{
\multirow{2}{*}{SDSS Object ID} & \colhead{R.A.} & \colhead{Decl.} & 
\multirow{2}{*}{redshift} & \colhead{$r_{petro, 0}$} & \colhead{$M_{r, K-cor}^{*}$} & \colhead{$r_{cModel, SDSS, 0}$} & \colhead{$r_{cModel, HSC, 0}$} & \multirow{2}{*}{$\dn$} & \multirow{2}{*}{$\log~(M_{*} / \Msun)$} & \multirow{2}{*}{z Source} \\
 & (deg) & (deg) & & (mag) & (mag) & (mag) & (mag) & & }
\startdata
1237661850937262105 & 200.000865 & 43.924572 & $0.201490 \pm 0.000095$ & $19.12 \pm 0.04$ & -20.52 & $19.03 \pm 0.02$ & $18.70 \pm 0.01$ & $1.47 \pm 0.06$ & $10.28 \pm ^{0.23}_{0.12}$ & MMT \\
1237661850400391774 & 200.000936 & 43.526926 & $0.197847 \pm 0.000229$ & $20.45 \pm 0.09$ & -19.05 & $20.40 \pm 0.05$ & $20.24 \pm 0.01$  & $1.51 \pm 0.09$ & $ 9.93 \pm ^{0.17}_{0.18}$ & MMT \\
1237662196141916862 & 200.001328 & 42.629571 & $0.481113 \pm 0.000100$ & $21.08 \pm 0.16$ & -21.08 & $21.06 \pm 0.07$ & $20.90 \pm 0.01$  & $1.28 \pm 0.07$ & $10.30 \pm ^{0.24}_{0.24}$ & MMT \\
1237661850937262450 & 200.001377 & 43.906918 & $0.394855 \pm 0.000217$ & $21.22 \pm 0.17$ & -20.28 & $21.23 \pm 0.07$ & $21.06 \pm 0.01$  & $1.90 \pm 0.28$ & $10.61 \pm ^{0.11}_{0.15}$ & MMT \\
1237661968508649886 & 200.001869 & 42.574824 & $0.359950 \pm 0.000095$ & $19.44 \pm 0.02$ & -21.75 & $19.43 \pm 0.02$ & $19.44 \pm 0.01$  & $1.75 \pm 0.03$ & $11.13 \pm ^{0.08}_{0.14}$ & MMT \\
1237661871871426772 & 200.002381 & 43.335269 & $0.230043 \pm 0.000105$ & $17.83 \pm 0.03$ & -21.98 & $17.89 \pm 0.01$ & $18.40 \pm 0.01$  & $1.77 \pm 0.04$ & $11.32 \pm ^{0.07}_{0.12}$ & MMT \\
1237661872408298189 & 200.002811 & 43.608926 & $0.360782 \pm 0.000162$ & $20.15 \pm 0.07$ & -21.14 & $20.03 \pm 0.03$ & $19.91 \pm 0.01$  & $1.49 \pm 0.08$ & $10.60 \pm ^{0.13}_{0.14}$ & MMT \\
1237661871334621428 & 200.007639 & 42.860764 & $0.181610 \pm 0.000062$ & $18.35 \pm 0.03$ & -20.97 & $18.20 \pm 0.01$ & $17.96 \pm 0.01$  & $1.75 \pm 0.04$ & $11.10 \pm ^{0.07}_{0.06}$ & MMT \\
1237661871334621440 & 200.007805 & 42.789304 & $0.228879 \pm 0.000145$ & $18.25 \pm 0.02$ & -21.76 & $18.21 \pm 0.01$ & $17.66 \pm 0.01$  & $1.55 \pm 0.03$ & $10.57 \pm ^{0.14}_{0.10}$ & MMT \\
1237661849863586400 & 200.007981 & 43.105365 & $0.503889 \pm 0.000104$ & $20.34 \pm 0.08$ & -22.12 & $20.28 \pm 0.05$ & $19.85 \pm 0.01$  & $1.55 \pm 0.10$ & $10.89 \pm ^{0.16}_{0.19}$ & MMT 
\enddata 
{\bf Notes.} The entire table is available in machine-readable form in the online journal. Here, a portion is shown for guidance regarding its format.
\tablenotetext{*}{K-correction at $z = 0.35$}
\end{deluxetable*}

\section{Survey Completeness} \label{sec:complete}

Survey completeness is a key aspect of a redshift survey. HectoMAP provides a uniform and complete spectroscopic survey for red-selected galaxies in the survey area. The complete HectoMAP survey enables, for example, the identification of galaxy clusters and investigation of quiescent galaxy evolution (e.g., \citealp{Sohn21b, Damjanov22b}). We evaluate the general HectoMAP survey completeness in Section \ref{sec:comp_general} and the HectoMAP survey completeness for quiescent galaxies in Section \ref{sec:comp_quiescent}. 

\subsection{HectoMAP Completeness} \label{sec:comp_general}

\begin{figure}
\centering
\includegraphics[scale=0.22]{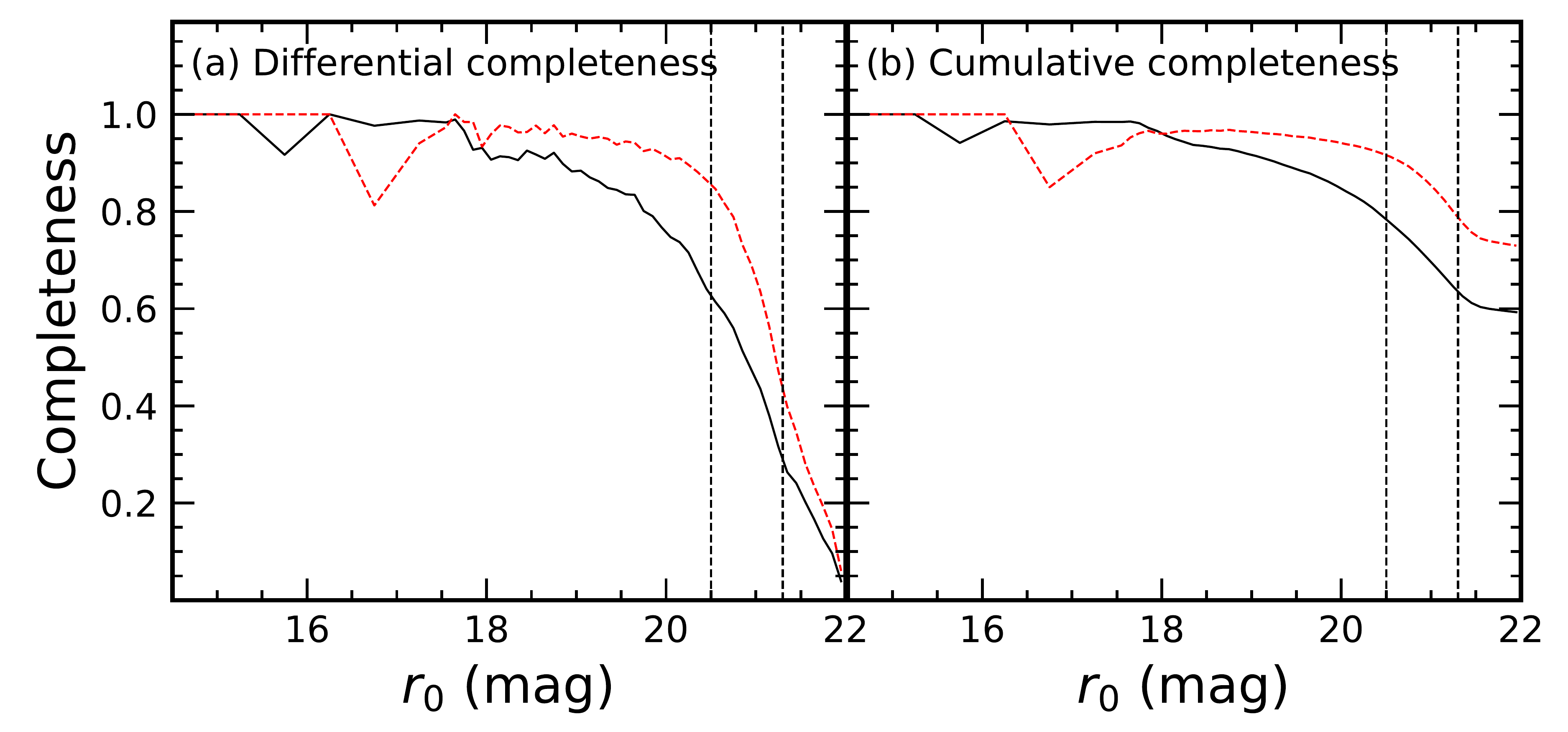} 
\caption{(a) Differential and (b) cumulative completeness of the HectoMAP spectroscopic survey as a function of $r_{cModel, 0}$. Black solid and red dashed lines show the completeness for galaxies with $(g-r) > 1.0$, and for galaxies with $(g-r) > 1.0$ and $(r-i) > 0.5$, respectively. }
\label{fig:completeness_cmd}
\end{figure}

Figure \ref{fig:completeness_cmd} show (a) the differential spectroscopic survey completeness and (b) the cumulative completeness as a function of $r-$band cModel magnitude. Although the original HectoMAP survey is based on Petrosian magnitudes, we use cModel magnitudes here for direct comparison with the completeness we compute based on HSC-SSP cModel magnitudes (Section \ref{sec:HSC}). In Figure \ref{fig:completeness_cmd}, the black solid line shows the completeness for the red-selected sample with $(g-r) > 1.0$. The red-selected sample is more than 80\% complete (the cumulative completeness) for $r \leq 20.5$ and 67\% complete for $r \leq 21.3$. With the additional survey color selection, $(r-i) > 0.5$ (red dashed line in Figure 9), HectoMAP is $\sim 95\%$ and $\sim 83\%$ for $r \leq 20.5$ and for $r \leq 21.3$, respectively.

\begin{figure*}
\centering
\includegraphics[scale=0.2]{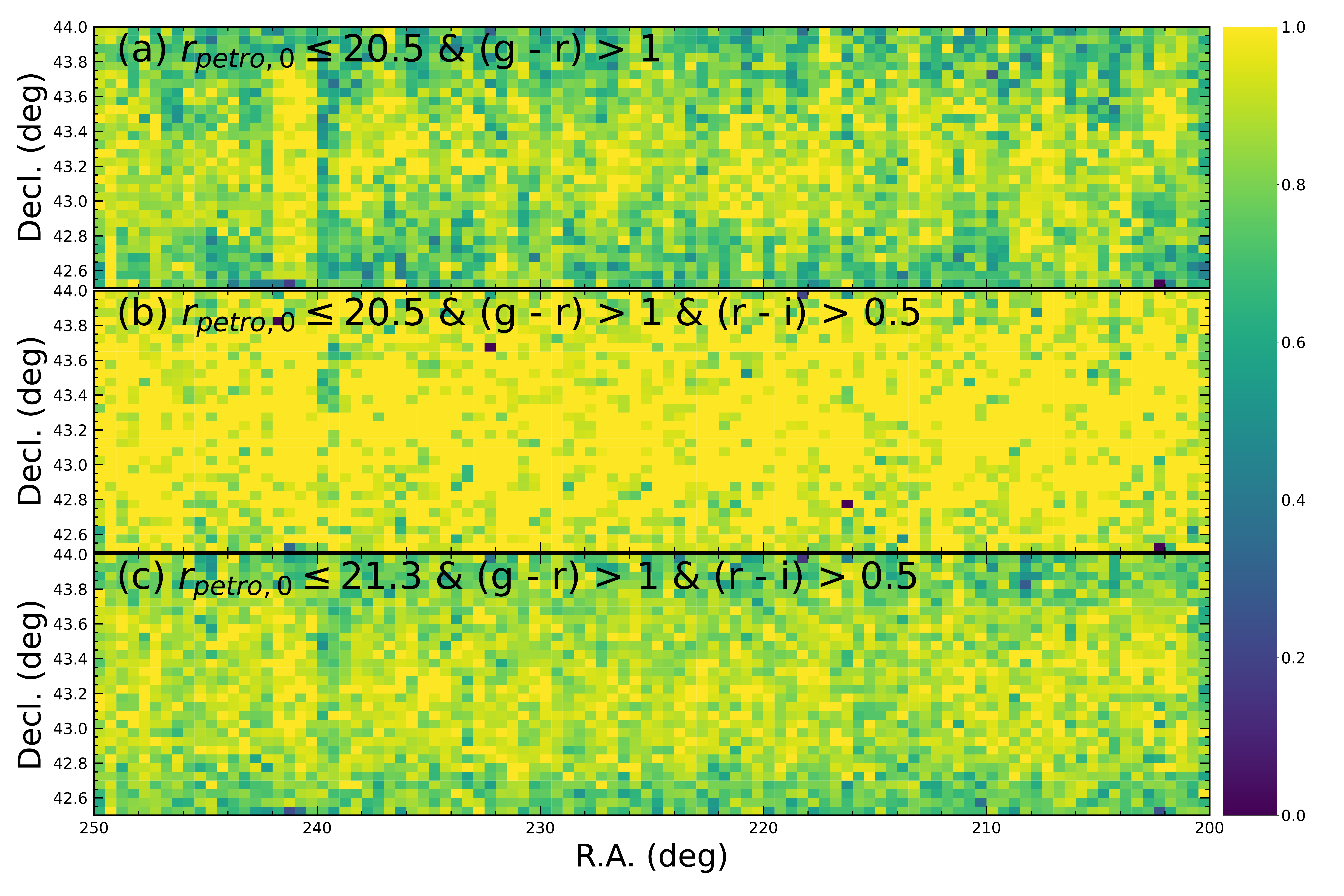} 
\caption{Completeness map for (a) bright galaxies with $r \leq 20.5$ and $(g-r) > 1.0$, (b) bright galaxies with $r \leq 20.5$, $(g-r) > 1.0$ and $(r-i) > 0.5$, and (c) faint galaxies with $r \leq 21.3$, $(g-r) > 1.0$, and $(r-i) > 0.5$. A lighter color indicates higher completeness. HectoMAP is uniformly complete over the survey area, but the completeness decreases toward the survey declination edges. }
\label{fig:completeness_map}
\end{figure*}

Figure \ref{fig:completeness_map} (a) displays the survey completeness map for galaxies brighter than $r = 20.5$ and redder than $(g-r) = 1.0$. We also display maps for red galaxies with $(g-r) > 1.0$ and $(r-i) > 0.5$ (Figure \ref{fig:completeness_map} (b)). Finally, we show the completeness map for the galaxies with $r \leq 21.3$, $(g-r) > 1.0$ and $(r-i) > 0.5$ (Figure \ref{fig:completeness_map} (c)). A lighter color indicates higher completeness. 

HectoMAP is uniformly complete over the entire survey field in general, but the completeness decreases toward the edges of the survey. For example, the completeness for a strip covering the central 0.05 degree declination area is $\sim 90\%$; a similar strip that covers the edge is $\sim 75\%$ complete. The survey is also slightly more complete for $240 < $ R.A. (deg) $ < 242$. This area is the GTO2deg$^{2}$ field \citep{Kurtz12} where we tested the feasibility of HectoMAP. Because observing conditions were better when we observed the GTO field, we were able to make 12 visits per unit area in the allocated time rather than the 9 average visits that characterize the full survey. 

\subsection{HectoMAP Quiescent Galaxy Completeness} \label{sec:comp_quiescent}

Because HectoMAP is red-selected, the survey is a robust platform for studying the quiescent galaxy population. For example, \citet{Damjanov22b} explore the size-mass relation of quiescent galaxies. They identify newcomers that join the quiescent population at $z \lesssim 0.5$. \citet{Damjanov22b} demonstrate that these newcomers typically have larger size at a given stellar mass than their older counterparts. The high completeness of the HectoMAP survey minimizes systematic biases that may originate from incompleteness. HectoMAP enables the definition of complete mass-limited subsamples of the survey. Here we highlight the completeness of the HectoMAP quiescent population.

Following previous work (e.g., \citealp{Vergani08, Zahid16, Sohn21a, Damjanov22a, Hamadouche22}), we identify quiescent galaxies based on the spectral indicator $\dn$. We identify $\dn > 1.5$ galaxies as the quiescent population \citep{Woods10, Damjanov22a}. Here we test the impact of the HectoMAP red-selection on the completeness of the quiescent population by comparing HectoMAP with SHELS/F2 \citep{Damjanov22b}.

The SHELS/F2 survey (hereafter F2, \citealp{Geller14}) provides a unique opportunity for testing the completeness of the HectoMAP quiescent sample. The F2 survey is an MMT/Hectospec survey of the F2 field, one of the Deep Lens Survey (DLS) regions; the field covers 3.98 deg$^{2}$. The Hectospec survey for F2 is a magnitude-limited survey with no color selection. The F2 survey is a complete magnitude-limited to DLS $R \leq 20.6$. We convert the basis photometry from DLS $R$ to SDSS $r$ for more direct comparison with HectoMAP \citep{Damjanov22a}. The final F2 survey is then complete ($> 95\%$) for $r = 20.75$. 

Because the F2 survey is a purely magnitude-limited survey, it includes every quiescent galaxy brighter than the magnitude limit regardless of color. We investigate the way the color selection for HectoMAP affects the completeness of the HectoMAP quiescent galaxy sample. Following \citet{Damjanov22b}, we examine the survey completeness for quiescent galaxies as a function of stellar mass. We also use four redshift subsamples with $z < 0.2$, $0.2 \leq z < 0.3$, $0.3 \leq z < 0.4$, and $0.4 \leq z < 0.6$. 

\begin{figure*}
\centering
\includegraphics[scale=0.32]{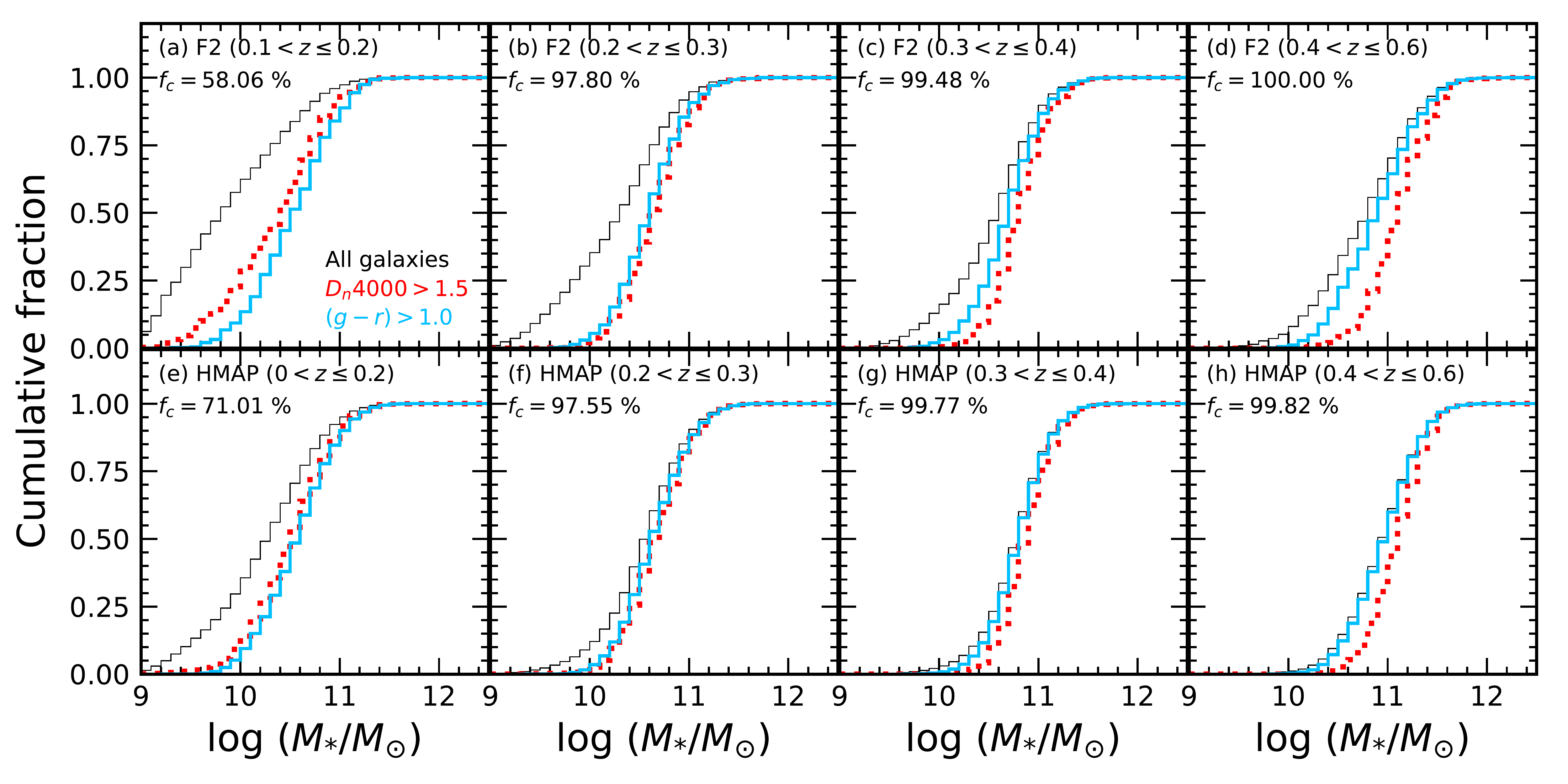} 
\caption{Cumulative fraction of galaxies with $r \leq 20.5$ as a function of stellar mass in four redshift bins: $0.1 < z \leq 0.2$, $0.2 < z \leq 0.3$, $0.3 < z \leq 0.4$, and $0.4 < z \leq 0.6$. The upper panels show SHELS/F2 galaxies; the lower panels display HectoMAP galaxies. Black solid lines show the cumulative distribution for all galaxies in each redshift bin. Red dotted lines display quiescent galaxies with $\dn > 1.5$ regardless of color. Blue solid lines show galaxies with $(g-r) > 1.0$. The numbers denote the recovery fraction for the quiescent population ($\dn > 1.5$) among the HectoMAP red ($g-r > 1.0$) galaxies. }
\label{fig:completeness_quiescent}
\end{figure*}

In Figure \ref{fig:completeness_quiescent}, black lines in the upper panels show the cumulative distribution for all F2 galaxies brighter than $r = 20.5$ as a function of stellar mass in the four different redshift bins. Red lines show F2 quiescent galaxies with $\dn > 1.5$. Because quiescent galaxies are generally more massive than their star-forming counterparts at the same apparent magnitude, the cumulative distributions of quiescent galaxies are shifted toward higher stellar mass.

We next apply the HectoMAP color selection, $(g-r) > 1.0$; we refer to this galaxy subsample as the red subsample. We emphasize that we do not apply a selection based on $\dn$ for the red subsample. The blue solid histograms display the cumulative stellar mass distribution of the red subsample. At low redshift $z < 0.2$, the red subsample includes generally higher stellar mass galaxies compared to the quiescent subsample. In other words, not all quiescent galaxies are included in the red subsample. In contrast, at higher redshift ($z > 0.2$), the cumulative stellar mass distributions of the red subsamples are very similar to those of the quiescent subsample.

We define the recovery fraction of quiescent galaxies ($\dn > 1.5$) with the $(g-r)$ color selection for the galaxies brighter than $r = 20.5$:
\begin{equation}
f_{recovery} = \frac{N_{red}((g-r) > 1.0)}{N_{quiescent} (\dn > 1.5)}.
\end{equation}
The recovery fraction for the $z < 0.2$ subsample is only 58\%. As shown in Figure \ref{fig:completeness_quiescent} (a), there are many low mass quiescent galaxies bluer than $(g-r) = 1.0$ in this redshift range. In contrast, the recovery fraction is remarkably high (97 - 100\%) at higher redshift. In other words, the red subsamples at $z > 0.2$ are complete for quiescent galaxies. This test substantiates the completeness of quiescent subsamples based on HectoMAP.

The lower panels of Figure \ref{fig:completeness_quiescent} show the cumulative stellar mass distributions of the HectoMAP subsamples. The HectoMAP subsamples are similar to the F2 subsamples especially for $z > 0.2$. 

\citet{Damjanov22b} carry out a similar test based on fainter galaxies ($r > 20.5$) with additional $(r-i) > 0.5$ selection (their Figure 2). They used the F2 fainter galaxies with $20.5 < r < 20.75$, the effective F2 survey completeness limit. They demonstrate that the $(r-i)$ color selection for fainter galaxies also recovers ($> 98\%$) the quiescent population (i.e., $\dn > 1.5$).

\section{Comparison between SDSS and HSC Photometry} \label{sec:HSC}

The HSC-SSP program completely covers the HectoMAP region \citep{Miyazaki12, Aihara22}. Because the HSC-SSP photometry is much deeper than the SDSS, we compare the SDSS with HSC-SSP as a basis for future applications of HectoMAP. 

We describe the HSC-SSP photometry in Section \ref{sec:HSCphot}. The cross-match of SDSS with HSC-SSP based on R.A. and Decl. identifies HSC counterparts for 92\% of the SDSS galaxies brighter than $r = 21.3$. HectoMAP galaxies without HSC counterparts are mostly near bright stars that are saturated in the HSC images and thus contaminate or even prevent photometry of the neighboring galaxies. Here, our analysis is based on the cModel magnitudes that are available in both SDSS and HSC-SSP photometry. 

\begin{figure}
\centering
\includegraphics[scale=0.22]{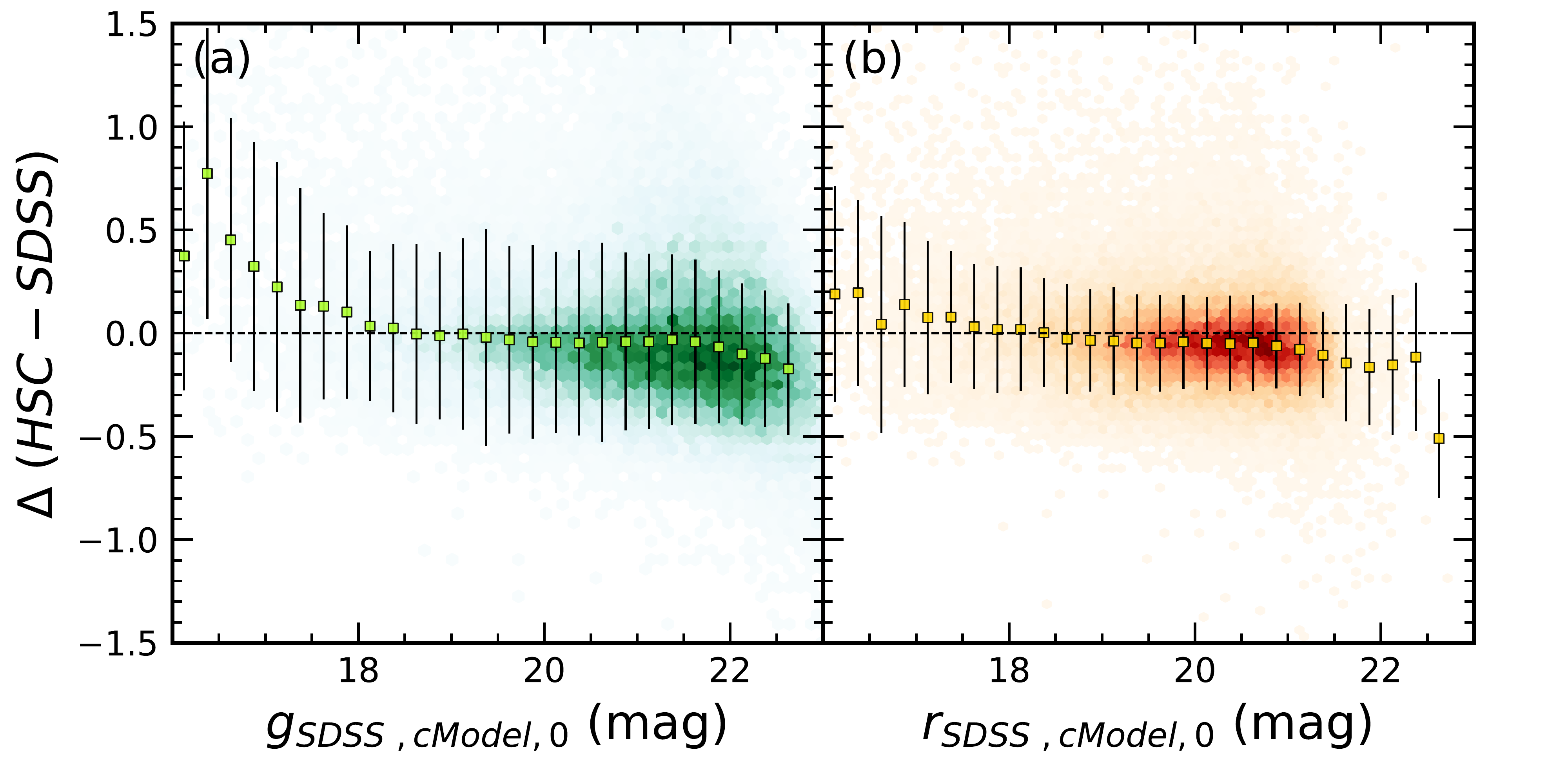}
\caption{(a) The $g-$band cModel magnitude difference between HSC and SDSS photometry as a function of the SDSS $g-$band magnitude. Large squares and error bars indicate the median difference and the $1\sigma$ standard deviation, respectively. (b) Same as (a), but for $r-$band cModel magnitudes. }
\label{fig:hsc_comp_mag}
\end{figure}

Figure \ref{fig:hsc_comp_mag} compares $g-$ and $r-$band cModel magnitudes from SDSS and HSC. We use only HectoMAP galaxies with spectroscopy for this comparison. The background density maps show the magnitude difference as a function of $g_{cModel, SDSS}$ or $r_{cModel, SDSS}$. The large squares indicate the median difference in each magnitude bin. 

The HSC photometry is generally fainter (magnitudes are larger) than the SDSS photometry for bright galaxies with $r < 19$. \citet{Bosch18} and \citet{Huang18b} argue that the deblending technique applied in {\it hscPipe} tends to over-subtract fainter objects and/or extended features surrounding the target galaxies; thus the total flux for bright galaxies is underestimated. In contrast, we suspect that the fainter outskirts of a faint galaxy detected cleanly in the deep HSC images results in a brighter HSC magnitude. 

We measure the median systematic difference between SDSS and HSC photometry at $18 < g_{cModel} < 21$ and $18 < r_{cModel} < 21$. The typical difference is only $-0.04$ mag in the $g-$band and $-0.05$ mag in the $r-$band. These differences are smaller than the typical uncertainties in the HSC cModel photometry. \citet{Huang18a} demonstrate that the typical uncertainty in cModel magnitudes for artificially injected galaxies are $\sim 0.15$ mag in both the $g-$ and $r-$ bands. Thus, the magnitude differences between SDSS and HSC-SSP photometry are well within the uncertainties of HSC-SSP photometry.  

\begin{figure}
\centering
\includegraphics[scale=0.34]{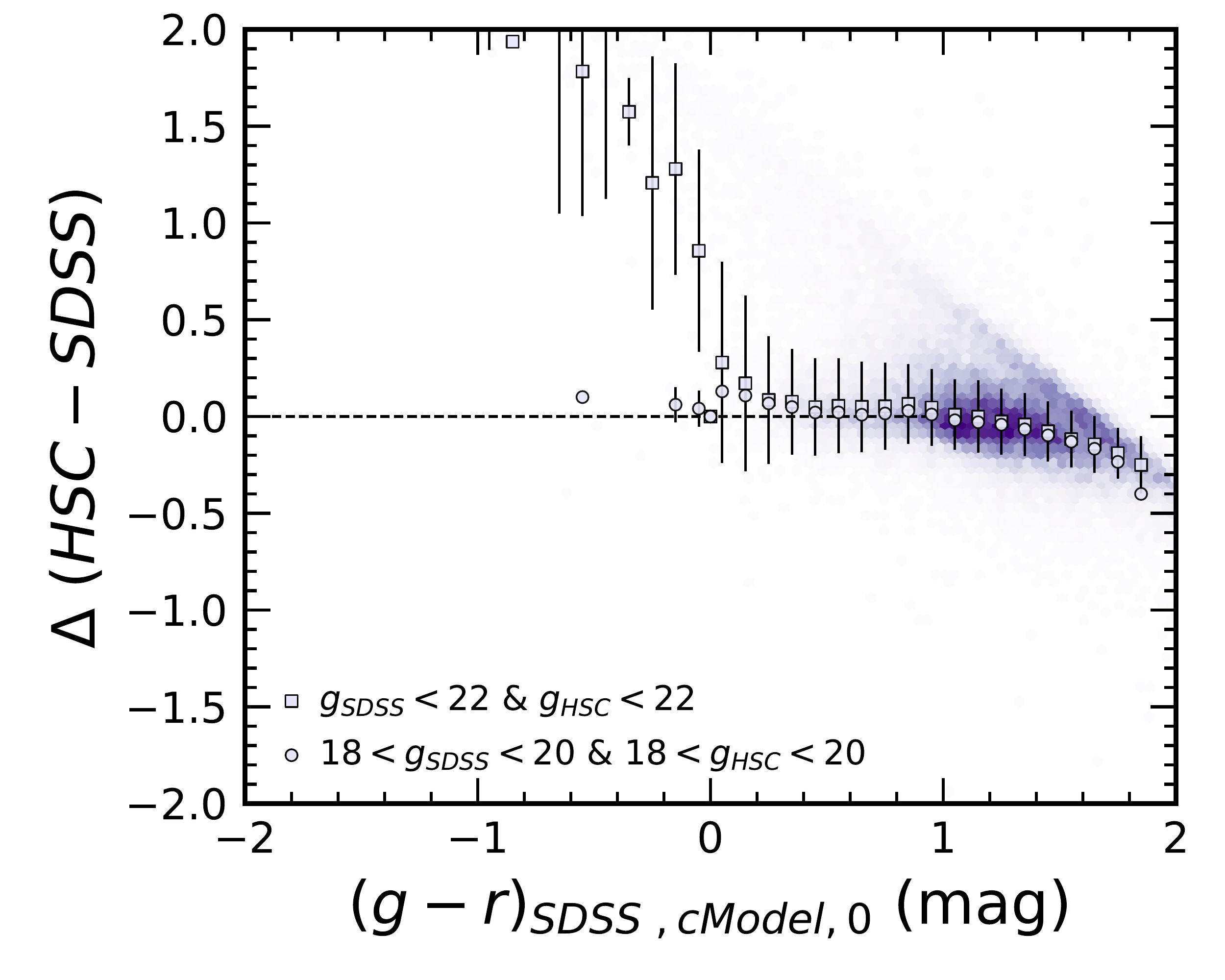}
\caption{Same as Figure \ref{fig:hsc_comp_mag}, but for $(g-r)_{cModel}$ color. }
\label{fig:hsc_comp_col}
\end{figure}

Figure \ref{fig:hsc_comp_col} compares the $(g-r)_{cModel}$ colors from SDSS and HSC. The definition of the symbols is the same as in Figure \ref{fig:hsc_comp_mag}. The color difference is significant for $(g-r)_{cModel} < 0.5$. These large differences result from significantly fainter HSC photometry in the $g-$band. Although the $(g-r)_{cModel}$ difference is significant for these blue $(g-r)$ colors, this difference does not affect our analyses because of the red-selection of HectoMAP. The circles in Figure \ref{fig:hsc_comp_col} display the median $(g-r)_{cModel}$ difference for galaxies with $18 < g_{SDSS, cModel, 0} < 20$ and $18 < g_{HSC, cModel, 0} < 20$. For these objects, the median $(g-r)_{cModel}$ difference is close to zero over a large color range.

\begin{figure}
\centering
\includegraphics[scale=0.22]{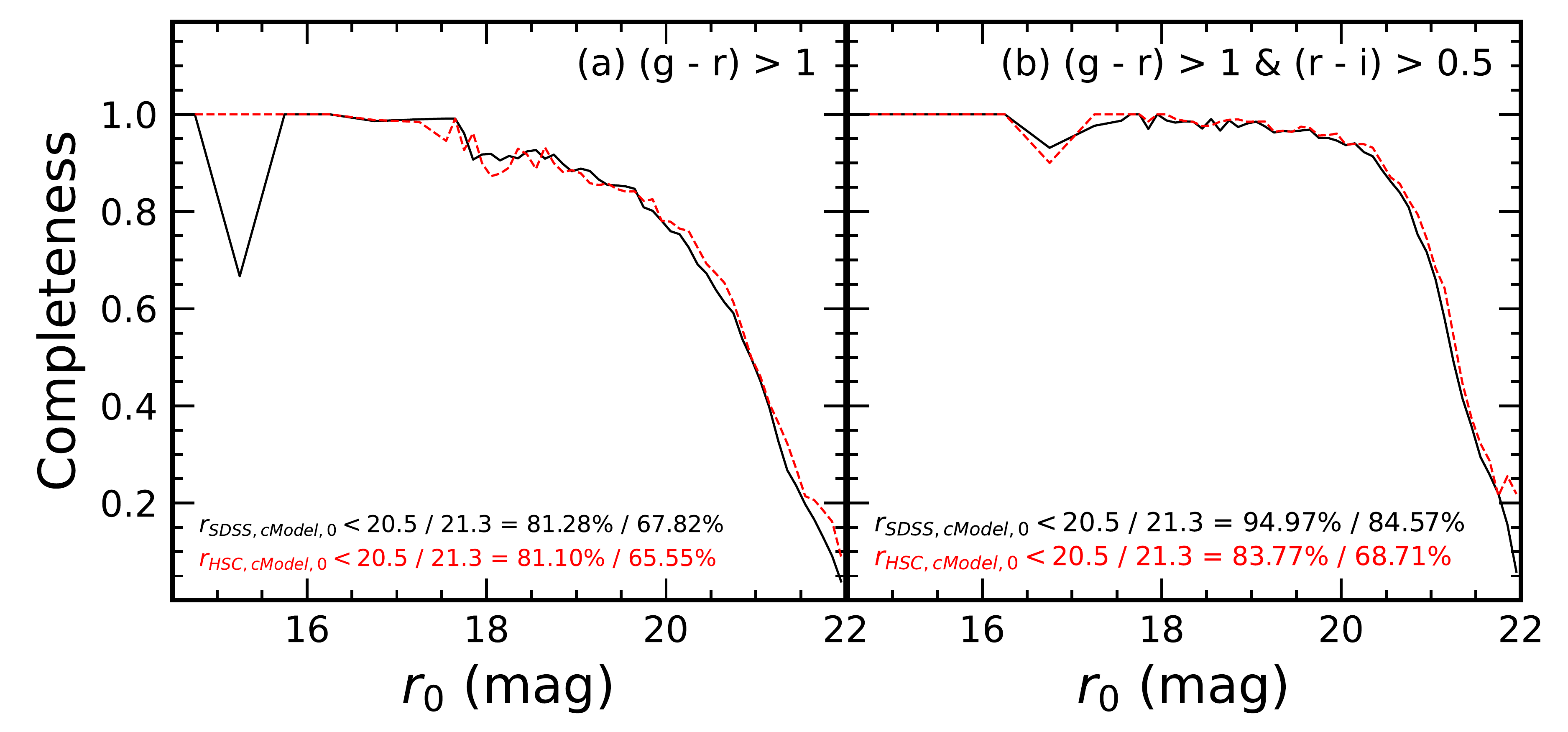}
\caption{(a)HectoMAP survey completeness for the galaxies with $(g-r)_{cModel} > 1.0$ based on the HSC and SDSS photometry. The black solid and red dashed lines show the completeness computed based on SDSS and HSC photometry, respectively. (b) Same as (a), but for galaxies with $(g-r)_{cModel} > 1.0$ and $(r-i)_{cModel} > 0.5$.}
\label{fig:hsc_zcomp}
\end{figure}

The spectroscopic survey completeness based on the HSC-SSP photometry (Figure \ref{fig:hsc_zcomp}) is an interesting aspect of HectoMAP. In fact, the completeness based on HSC photometry is remarkably consistent with that based on SDSS photometry: $\sim 81\%$ at $r_{cModel} < 20.5$ and $\sim 65\%$ at $r_{cModel} < 21.3$ for $(g-r)_{cModel} > 1.0$ galaxies. This consistency is a coincidence. Because of statistical errors in the photometry, particularly in the color, the SDSS and HSC samples are not identical. For example, there are 42,999 spectroscopic sample of galaxies with $r_{SDSS, cModel} < 20.5$ and $(g-r)_{SDSS, cModel} > 1.0$, but only 80\% of these objects have $r_{HSC, cModel} < 20.5$ and $(g-r)_{HSC, cModel} > 1.0$. Among the missing 8935 galaxies, 74\% of them are brighter than $r_{HSC, cModel} = 20.5$, but the $(g-r)_{HSC, cModel}$ colors are bluer than the limit. The relatively large uncertainty in the HSC $g-$band photometry underlies in the differences in the samples. Similarly, among 18,744 spectroscopic sample of galaxies with $20.5 \leq r_{SDSS, cModel} < 21.3$, $(g-r)_{SDSS, cModel} > 1.0$ and $(r-i)_{SDSS, cModel} > 0.5$, only 70\% (12959) of the galaxies satisfy a similar magnitude and color selection based on HSC photometry.

\section{Comparison between Photometric and Spectroscopic Redshifts} \label{sec:zphot}

\begin{figure}
\centering
\includegraphics[scale=0.33]{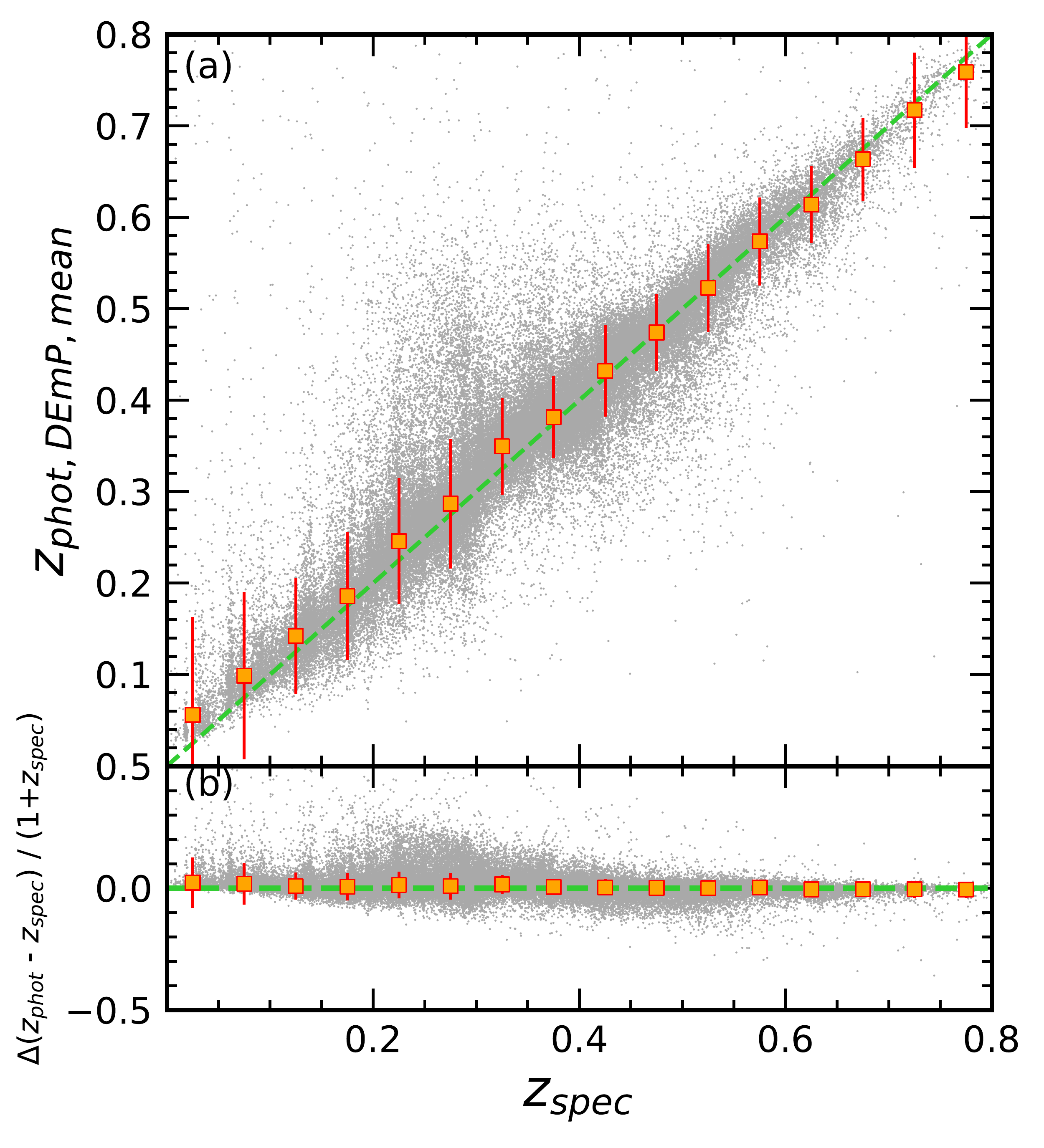} 
\caption{Comparison between HSC DEmP photometric and HectoMAP spectroscopic redshifts. Squares and error bars indicate the median and $1\sigma$ standard deviation as a function of spectroscopic redshift. }
\label{fig:zphot}
\end{figure}

The extensive HectoMAP spectroscopy provides a testbed for the updated photometric redshifts from HSC-SSP DR3 \citep{Aihara22}. In \citet{Sohn21a}, we compared HectoMAP DR1 spectroscopic redshifts with the photometric redshifts based on HSC-SSP DR2 photometry (hereafter DR2 $z_{phot}$). \citet{Sohn21a} used 17,040 HectoMAP DR1 objects with both spectroscopic and photometric redshifts. They showed that HSC photometric redshifts ($z_{phot}$) are generally consistent with spectroscopic redshifts ($z_{spec}$), but with large scatter. The difference between the $z_{phot}$ and $z_{spec}$ depended significantly on the apparent $i$-band magnitude.

Here we explore the $z_{phot}$ derived with the DEmP template fitting code \citep{Hsieh14} based on  HSC-SSP DR3 \citep{Aihara22}. \citet{Aihara22} emphasize  changes in the sky subtraction improved the DR3 photometry relative to DR2. These improvements modify the resulting photometric redshifts. We obtain {\it PHOTOZ\_MEDIAN}, the median value of the photometric redshift probability distribution function, from the DR3 DEmP catalog. 

Figure \ref{fig:zphot} compares $z_{spec}$ and $z_{phot}$ for 88,450 galaxies within the entire HectoMAP region. Figure \ref{fig:zphot} (a) displays a direct comparison between $z_{spec}$ and $z_{phot}$ and Figure \ref{fig:zphot} (b) shows the difference between $z_{spec}$ and $z_{phot}$ normalized by $(1 + z_{spec})$ as a function of $z_{spec}$. Red squares and error bars indicate the median and $1\sigma$ standard deviation of $z_{phot}$ and $(z_{phot} - z_{spec}) / (1+z_{spec})$ as a function of $z_{spec}$.

The typical difference between $z_{spec}$ and $z_{phot}$ in the spectroscopic redshift range $z_{spec} < 0.8$ is $0.006 \pm 0.050$ (i.e., $1800 \pm 15000~\kms$). This difference is comparable with the line-of-sight velocity dispersion of massive galaxy clusters. The typical uncertainties are much larger than the cluster velocity dispersion. The large scatter in $z_{phot}$ measurements thus significantly limits the power of photometric redshifts for studies of galaxy systems and the details of large-scale structure. For example, Figure 18 of \citet{Sohn21a} shows that the large-scale structure traced by $z_{phot}$ in the HectoMAP DR1 region is almost completely blurred by the scatter in the photometric relative to the spectroscopic redshift. 

For further examination of the HSC-SSP DR3 photometric redshifts in comparison with HectoMAP, we revisit three metrics that test the $z_{phot}$ algorithm following \citet{Tanaka18}: the bias, the conventional dispersion, and the loss function. The bias parameter ($\Delta z$) indicates the systematic offset between $z_{phot}$ and $z_{spec}$:
\begin{equation}
\Delta z = (z_{phot} - z_{spec}) / (1 + z_{spec}).
\end{equation}
The conventional dispersion measures the spread in the difference between $z_{phot}$ and $z_{spec}$. We compute the conventional dispersion $\sigma_{conv}$: 
\begin{equation}
\sigma_{conv} = 1.48 \times MAD(z),
\end{equation}
where $MAD(z)$ is the median absolute deviation. 
Finally, we compute the loss function defined by \citet{Tanaka18}:
\begin{equation}
L(\Delta z) = 1 - {1 \over{ (1 + ({{\Delta{z}}\over{\gamma}})^{2})}}, 
\end{equation}
where $\gamma = 0.15$ as in \citet{Tanaka18}. This loss function is a continuous version of the outlier fraction that includes both the bias and the dispersion. We follow \citet{Tanaka18} in taking $\gamma = 0.15$ that corresponds to the standard limit $\Delta{z} = 0.15$ used for computing the outlier fraction.

\begin{figure*}
\centering
\includegraphics[scale=0.35]{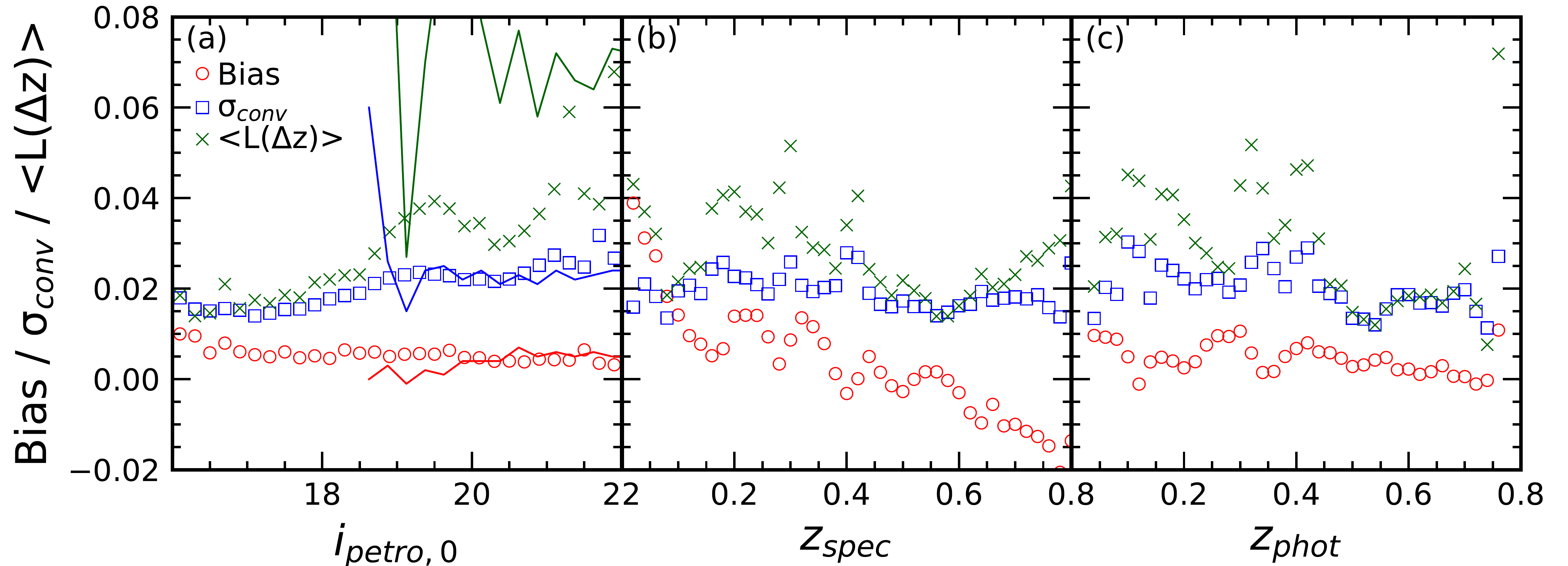} 
\caption{The bias ($\Delta z$, red circles), the conventional dispersion ($\sigma_{conv}$, blue squares), and the loss function ($<L (\Delta z)>$, green crosses) as a function of (a) SDSS $i-$band magnitude, (b) $z_{spec}$ and (c) $z_{phot}$. The solid lines in (a) show the distribution of the three metrics derived based on HSC-SSP PDR1 $z_{phot}$ \citep{Tanaka18}.}
\label{fig:zphot_bias}
\end{figure*}

Figure \ref{fig:zphot_bias} (a) shows the three metrics as a function of $i-$band magnitude. Red circles, blue squares, and green crosses indicate $\Delta z$, $\sigma_{conv}$, and the $<L (\Delta z)>$, respectively. The bias remains remains almost constant at $\sim 1600~\kms$ and has little variation with apparent magnitude. Both $\sigma_{conv}$ and $<L (\Delta z)>$ increase slightly at fainter magnitudes. 

The solid lines in Figure \ref{fig:zphot_bias} show the three metrics measured from the earlier HSC-SSP DR1 \citet{Tanaka18} as a function of magnitude. These metrics are consistent with those for on  HSC-SSP DR2 $z_{phot}$  (Figure 19, \citet{Sohn21a}). Figure \ref{fig:zphot_bias} shows that the bias and $\sigma_{conv}$ measured from DR2 and DR3 $z_{phot}$ are essentially the same. However, the $<L (\Delta z)>$ measured from DR3 $z_{phot}$ is significantly smaller at $i < 21.5$ than it was for DR2. In other words, the DR3 photometric redshifts $z_{phot}$ are a significant improvement over DR2. The revised sky subtraction procedures \citep{Aihara22} are probably responsible at least in part for this improvement.

Figure \ref{fig:zphot_bias} (b) and (c) display the same metrics as a function of $z_{spec}$ and $z_{phot}$, respectively (See Figure 19 of \citet{Sohn21a} for the DR2 comparison). All three metrics show similar trends with either $z_{spec}$ or $z_{phot}$. The bias decreases slightly as a function of spectroscopic redshift. The value of the bias is  zero at $z \sim 0.4$; the bias is positive (negative) at lower (higher) redshift. This trend differs from the bias based on DR2 $z_{phot}$, where the bias is positive over the entire redshift range we explore  \citep{Sohn21a}. The $\sigma_{conv}$ changes little as a function of redshift as for DR2 photometric redshifts \citep{Sohn21a}.  

Interestingly, the loss function shows a distinctive behavior compared to that of the DR2 $z_{phot}$ \citep{Sohn21a}. For DR2 $z_{phot}$, the loss function is larger than 0.1 at $z < 0.1$ and remains constant at $\sim 0.06$ at $z > 0.1$. In contrast, the DR3 $z_{phot}$ loss function is generally below the DR2 result at every redshift. There are significant fluctuations in the loss function in the range $0.2 < z < 0.4$ (Figure \ref{fig:zphot}) both as a function of $z_{spec}$ and $z_{phot}$. This effect probably results from the larger number of well-sampled many galaxy systems in this redshift range. Naturally, larger photometric uncertainties in these crowded fields can affect the $z_{phot}$ measurements.

\section{Applications of HectoMAP to Clusters and the Cosmic Web}\label{sec:apps}

The area of HectoMAP (54.64 deg$^{2}$) combined with its density ($\sim 1800$ galaxies deg$^{-2}$) and depth enable exploration of a wide range of scientific issues. We summarize some of the existing projects enabled by the survey and we preview a few of its future applications.

The high density of the red-selected HectoMAP survey makes it a foundation for the study of clusters of galaxies in the redshift range $0.2 < z < 0.6$ \citep{Sohn18a, Sohn18b, Sohn21b}. There are 104 photometrically selected redMaPPer clusters in the HectoMAP region \citep{Rykoff16, Sohn18a}. More than 90\% of redMaPPer clusters include 10 or more HectoMAP spectroscopic members, a basis for evaluating the redMaPPer membership probability and for defining a spectroscopic richness. 

\citet{Sohn21b} identify galaxy clusters based purely on spectroscopy by applying a friends-of-friends algorithm. There are 346 galaxy overdensities with 10 or more spectroscopic members. Even among relatively dense systems in this friends-of-friends catalog, $\sim 70\%$ have no redMaPPer counterpart suggesting that the solely photometric catalog may be incomplete.

HectoMAP is the first survey that traces cluster infall regions, the link between clusters and the cosmic web. HectoMAP traces the infall regions for clusters in the mass range $M_{200} > 5 \times 10^{13}~\Msun$ for the redshift range $0.2 < z < 0.6$. \citet{Pizzardo22} use a technique developed by \citet{deBoni16} to measure the growth rate of clusters directly over the HectoMAP redshift range. This technique, which depends on observations of the infall region, yields a growth rate consistent with predictions of simulations.

One of the special strengths of HectoMAP is the combination of the dense spectroscopy with the deep HSC-SSP imaging. This combination enables a range of projects that combine two of the most powerful cosmological tools, lensing and redshift surveys.

Shapes for weak lensing sources have not yet been released for the entire HectoMAP region. Eventually, these data will enable weak lensing studies ranging from examining of the masses of quiescent galaxies and their evolution to tracing the matter distribution in the cosmic web by comparing the foreground galaxy distribution with the weak lensing map (e.g., \citealp{Utsumi16}). For the friends-of-friends clusters, weak lensing adds another mass proxy. 

\begin{figure*}
\centering
\includegraphics[scale=0.75]{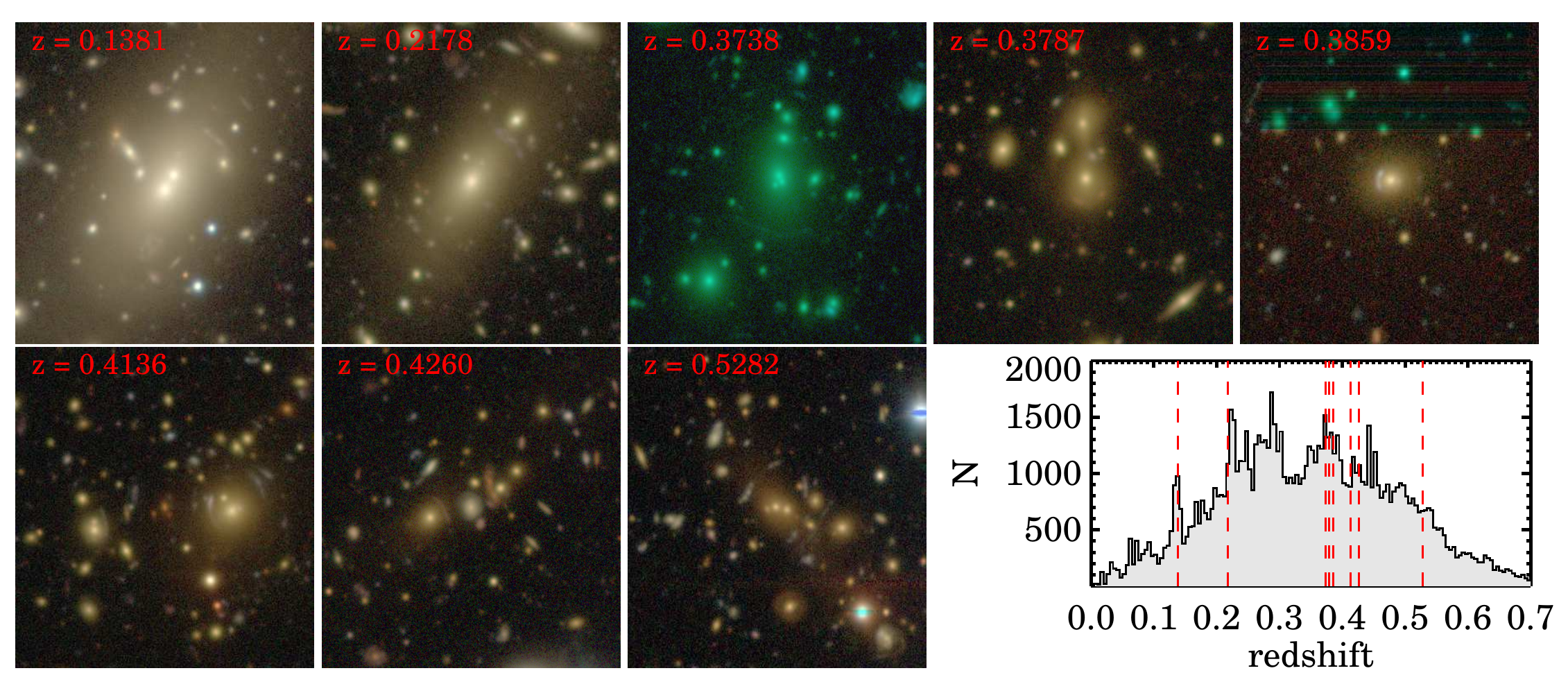} 
\caption{HSC images of 8 strong lensing clusters identified by both \citet{Jaelani20} and \citet{Sohn21b}. The lower inset shows the redshift distribution of HectoMAP galaxies (black histogram) and the redshifts of the strong lensing clusters (red vertical lines). }
\label{fig:slens}
\end{figure*}

\citet{Jaelani20} identify 23 candidate strong lensing systems in the HectoMAP region from HSC-SSP. Many of these systems have redshifts beyond the HectoMAP effective limit, but 8 of these systems match friends-of-friends systems. Figure \ref{fig:slens} shows HSC images of the 8 strong lensing systems. The lower right panel of Figure \ref{fig:slens} shows the redshift distribution of the strong lensing systems compared with the HectoMAP redshift distribution. In general, the more abundant, higher redshift systems have more obvious substructures frequently with multiple BCGs. Although this sample is small, it promises probes of cluster evolution, the relation between clusters and their BCGs, and the underlying cluster geometry that produces a strong lensing system.
 
HectoMAP densely samples the filaments and walls in the cosmic web that delineate the void regions. \citet{Hwang16} identify the underdense void regions in HectoMAP at $z > 0.2$. They show that the observed void properties including size and volume are in agreement with predicted void properties based on numerical simulations. \citet{Geller10} used the SHELS survey and DLS photometry to pioneer the direct comparison between structure in a redshift survey and a weak lensing map. A similar comparison based on HectoMAP and HSC-SSP will probe the relative distribution of light-emitting and dark matter on large scales.

HectoMAP is a testbed of approaches to understanding the mass distribution in the universe and its evolution. Future large, deeper surveys including DESI \citep{Zhou22}, MOONS \citep{Taylor18}, WEAVE \citep{Dalton12}, 4MOST \citep{Richard19}, Subaru/Prime Focus Spectrograph (PFS, \citealp{Takada14, Greene22}) survey will enhance these approaches and develop techniques taking advantage of the large extent and depth of these surveys.

\section{CONCLUSION} \label{sec:conclusion}

HectoMAP is a dense redshift survey with a median redshift of $z =0.345$. It densely covers 54.64 square degrees in a 1.5 degree wide strip across the northern sky centered at a declination of 43.25$^\circ$. The survey is dense ($\sim 1800$ galaxies deg$^{-2}$) and focused on the quiescent galaxy population. The survey includes 95,403 galaxies: among these galaxies, 60,237 are quiescent.

On average there were 9 visits to each position in the survey. Thus the overall completeness is high and uniform. The completeness is $\sim 81\%$ and $\sim 72\%$ for galaxies with $r < 20.5$ and $(g-r) > 1.0$, and with $20.5 < r < 21.3$, $(g-r) > 1.0$, and $(r-i) > 0.5$, respectively. 
 
\cite{Sohn21a} published a first data release covering 8.7 deg$^{2}$ of the survey area. They show individual spectra and substantiate the choice of a limiting cross-correlation value to identify objects with reliable redshifts. They also make a detailed comparison between HectoMAP and SDSS/BOSS redshifts. The small offset of $26~\kms$ and $46~\kms$ persists for the larger sample. These small offsets are comparable with the redshift errors and do not affect any of the analysis carried out so far with HectoMAP data.

In addition to the 88294 MMT/Hectospec redshifts provided here, we  include the spectral diagnostic ($\dn$) and stellar masses for 88165 galaxies. We highlight the distributions of these quantities in the HectoMAP survey data. As expected for a red-selected survey, the majority of the objects (63\%) are quiescent. The distribution of stellar masses is skewed toward large stellar mass reflecting this selection.

As \citet{Damjanov22b} emphasize, HectoMAP provides one of the largest, complete, mass-limited samples of quiescent galaxies covering the redshift range $0.2 < z < 0.6$. The sample includes 30,231 galaxies. \citet{Damjanov22b} use the sample to trace the size-mass relation of the quiescent population and to track the growth of various subpopulations with cosmic time.

One of the distinctive characteristics of HectoMAP is the full HSC-SSP coverage of the entire region. Observations for HectoMAP occurred over a 10 year period and thus the photometric basis was limited to the SDSS. We compare the SDSS DR16 photometry with the HSC-SSP photometry and show that the typical differences between the SDSS and HSC-SSP $g-$ and $r-$ band photometry are small compared with the photometric error. We also show that for red galaxies that are the bulk of the HectoMAP sample, the $(g-r)$ colors also agree; the difference for blue objects is significant. Remarkably, the completeness of HectoMAP is similar to both the SDSS and HSC-SSP magnitude limits, but the underlying samples in the two cases differ as a result of photometric error. 

The extensive HectoMAP spectroscopic redshifts provide a platform for testing HSC-SSP photometric redshifts. We revisit the comparison between $z_{phot}$ and $z_{spec}$ reported in \citet{Sohn21a}. Here we used the updated HSC-SSP DR3 $z_{phot}$. The DR3 $z_{phot}$ measurements are a significant improvement over the earlier DR2 photometric redshifts. In general, $z_{phot}$ agrees well with $z_{spec}$, but the scatter remains large ($\sim 1800 \pm 15,000~\kms$) compared with the typical velocity dispersions of galaxy systems.

Finally we highlight the particular power of HectoMAP for exploring clusters of galaxies. The survey contains an independently identified set of strong lensing clusters in the redshift range $0.2 < z < 0.6$. When full HSC-SSP shapes are available, weak lensing mass profiles combined with extensive on the strong lensing systems will elucidate evolutionary issues and the role of substructure along the line-of-sight in enhancing the lensing cross-section. More generally, HectoMAP covers a redshift range where clusters of galaxies accrete roughly half of their mass (e.g., \citealp{Fakhouri10, Sohn22}). \citet{Pizzardo22} measure the accretions rate for HectoMAP clusters and show that it agrees with predictions. Taken together these explorations of cluster growth and structure will inform the design of future larger, deeper dense surveys.

At the moment, HectoMAP has some distinctive features in the universe of redshift surveys; it is dense and red-selected. For example, in contrast with the 1800 galaxy deg$^{-2}$ density of HectoMAP, the DESI bright galaxy sample \citep{RuizMacias21}, though it will be much larger, reaches a limiting $r < 19.5$ with only 864 galaxies deg$^{-2}$ and the BGS faint sample covers the range $19.5 < r < 20.175$ with a density of 533 galaxies deg$^{-2}$. HectoMAP is a platform for investigating issues including the nature and evolution of the quiescent population, the growth of clusters of galaxies, and the way galaxies trace the matter distribution in the universe. The HectoMAP data and analyses will contribute to informing the design and goals of ongoing major surveys.

\begin{acknowledgments}

We first thank Perry Berlind and Micheal Calkins for operating Hectospec. We thank Susan Tokarz, Jaehyon Rhee, Sean Moran, Warren Brown, and Nelson Caldwell for their tremendous contributions to the observing preparation and  data reduction. We also thank Antonaldo Diaferio, Ken Rines, Ian Dell'Antonio, and Scott Kenyon for insightful discussions that clarified the paper. We thank the referee for suggesting a comparison with the updated HSC-SSP photometric redshifts. This work was supported by the New Faculty Startup Fund from Seoul National University. J.S. is also supported by a CfA Fellowship. M.J.G. and D.G.F acknowledge the Smithsonian Institution for support. H.S.H. acknowledges support by the National Research Foundation of Korea (NRF) grant funded by the Korean government (MSIT) (No. 2021R1A2C1094577). I.D. acknowledges the support of the Canada Research Chair Program and the Natural Sciences and Engineering Research Council of Canada (NSERC, funding reference number RGPIN-2018-05425).

Funding for the Sloan Digital Sky Survey IV has been provided by the Alfred P. Sloan Foundation, the U.S. Department of Energy Office of Science, and the Participating Institutions. SDSS-IV acknowledges support and resources from the Center for High Performance Computing at the University of Utah. The SDSS website is www.sdss.org. SDSS-IV is managed by the Astrophysical Research Consortium for the Participating Institutions of the SDSS Collaboration including the Brazilian Participation Group, the Carnegie Institution for Science, Carnegie Mellon University, Center for Astrophysics | Harvard \& Smithsonian, the Chilean Participation Group, the French Participation Group, Instituto de Astrof\'isica de Canarias, The Johns Hopkins University, Kavli Institute for the Physics and Mathematics of the Universe (IPMU) / University of Tokyo, the Korean Participation Group, Lawrence Berkeley National Laboratory, Leibniz Institut f\"ur Astrophysik Potsdam (AIP), Max-Planck-Institut f\"ur Astronomie (MPIA Heidelberg), Max-Planck-Institut f\"ur Astrophysik (MPA Garching), Max-Planck-Institut f\"ur Extraterrestrische Physik (MPE), National Astronomical Observatories of China, New Mexico State University, New York University, University of Notre Dame, Observat\'ario Nacional / MCTI, The Ohio State University, Pennsylvania State University, Shanghai Astronomical Observatory, United Kingdom Participation Group, Universidad Nacional Aut\'onoma de M\'exico, University of Arizona, University of Colorado Boulder, University of Oxford, University of Portsmouth, University of Utah, University of Virginia, University of Washington, University of Wisconsin, Vanderbilt University, and Yale University.

The Hyper Suprime-Cam (HSC) collaboration includes the astronomical communities of Japan and Taiwan, and Princeton University. The HSC instrumentation and software were developed by the National Astronomical Observatory of Japan (NAOJ), the Kavli Institute for the Physics and Mathematics of the Universe (Kavli IPMU), the University of Tokyo, the High Energy Accelerator Research Organization (KEK), the Academia Sinica Institute for Astronomy and Astrophysics in Taiwan (ASIAA), and Princeton University. Funding was contributed by the FIRST program from the Japanese Cabinet Office, the Ministry of Education, Culture, Sports, Science and Technology (MEXT), the Japan Society for the Promotion of Science (JSPS), Japan Science and Technology Agency (JST), the Toray Science Foundation, NAOJ, Kavli IPMU, KEK, ASIAA, and Princeton University. This paper makes use of software developed for the Large Synoptic Survey Telescope. We thank the LSST Project for making their code available as free software at http://dm.lsst.org. This paper is based [in part] on data collected at the Subaru Telescope and retrieved from the HSC data archive system, which is operated by Subaru Telescope and Astronomy Data Center (ADC) at National Astronomical Observatory of Japan. Data analysis was in part carried out with the cooperation of Center for Computational Astrophysics (CfCA), National Astronomical Observatory of Japan.
\end{acknowledgments}

\appendix \label{sec:appendix}

Table \ref{tab:star} lists the 6544 stars with spectroscopy. There are 4253 stars that were identified as point sources based in the SDSS database. Additionally, we spectroscopically identify 2291 stars originally identified as extended sources based in the SDSS database (hereafter, spectroscopically identified stars). We list SDSS Object ID, R.A., Decl. redshift and its uncertainty and $r-$band magnitude. We include only the SDSS photometry because many of stars are saturated in the HSC images. 

\begin{figure}
\centering
\includegraphics[scale=0.7]{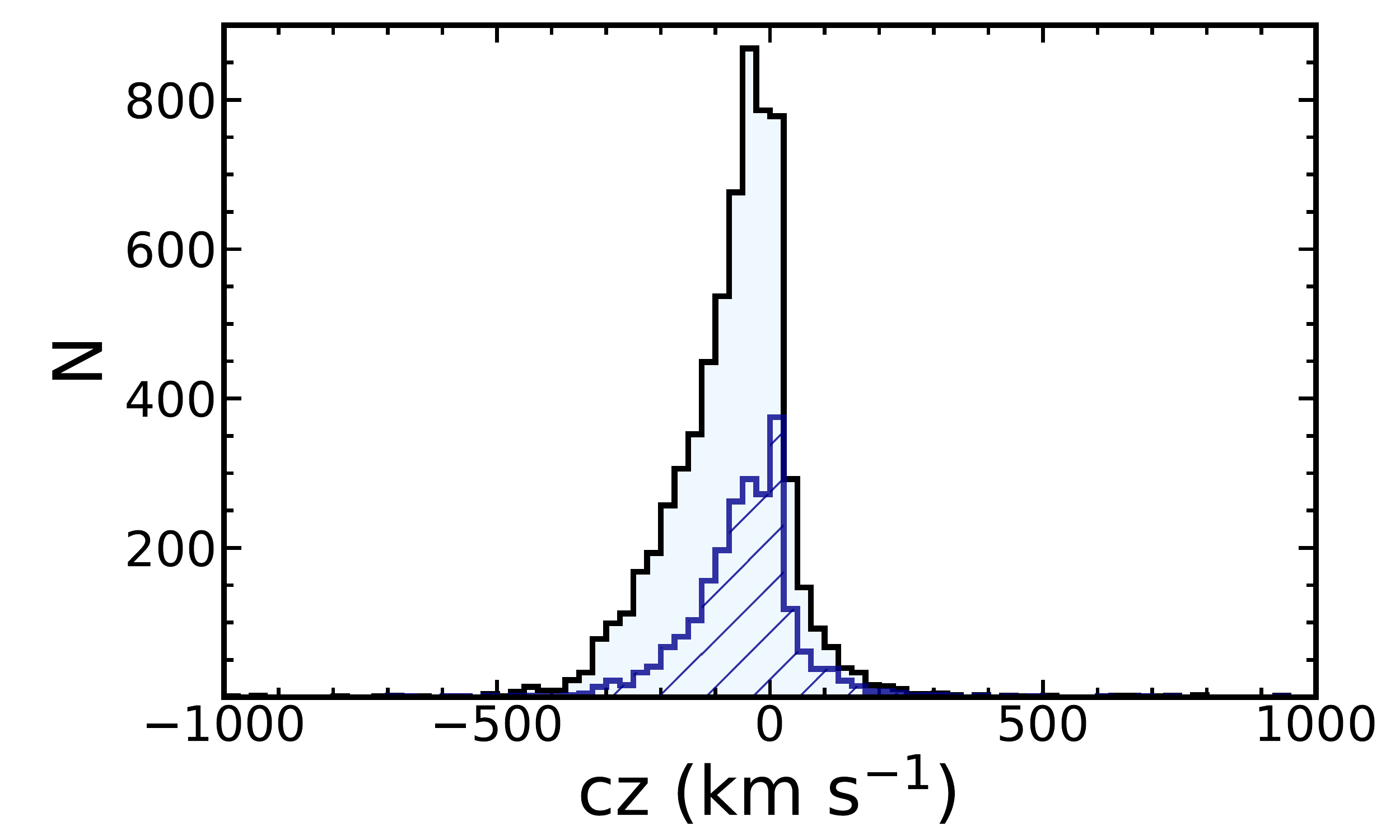}
\caption{Radial velocity distribution of stars with HectoMAP spectroscopy. The hatched histogram shows the distribution for stars identified as extended sources in the SDSS photometry alone.}
\label{fig:star}
\end{figure}

Figure \ref{fig:star} displays the radial velocity distribution of all of the stars in the HectoMAP region.  We mark the radial velocity distribution of extended objects in SDSS that are spectroscopically identified as stars with a hatched histogram. Because the sample of stars is not complete in any way, we include these data only as a guide for future surveys of the region.

\begin{deluxetable}{cccccc}
\label{tab:star}
\tablecaption{Radial Velocities for Stars in HectoMAP}
\tablecolumns{6}
\tablewidth{0pt}
\tablehead{
\multirow{2}{*}{SDSS Object ID} & \colhead{R.A.} & \colhead{Decl.} & 
\multirow{2}{*}{RV$^{a}$} & \colhead{$r_{petro, 0}$} & \colhead{Flag$^{b}$} \\
 & (deg) & (deg) & ($\kms$) & (mag) & }
\startdata
1237661360226500918 & 208.899206 &  43.773073 & $  108.82 \pm 122.02$ & $21.13 \pm 0.13$ & Extended \\
1237661360226500771 & 208.808559 &  43.794296 & $   -36.48 \pm   29.98$ & $20.49 \pm 0.07$ &        Point \\
1237661360763633780 & 209.840059 &  43.983869 & $   -40.41 \pm     6.18$ & $17.29 \pm 0.00$ &        Point \\
1237661433241862442 & 208.388763 &  43.964662 & $ -262.36 \pm   28.11$ & $20.90 \pm 0.06$ &        Point \\
1237661433242058831 & 208.999039 &  43.988056 & $   -19.59 \pm   16.66$ & $21.75 \pm 1.40$ &        Point \\
1237661433242190333 & 209.325341 &  43.897316 & $    18.45 \pm   11.00$ & $21.27 \pm 0.23$ &        Point \\
1237661433242189895 & 209.230960 &  43.803751 & $   -78.14 \pm     7.72$ & $18.85 \pm 0.01$ &        Point \\
1237661849863651845 & 200.241813 &  43.048129 & $   -76.82 \pm   13.45$ & $20.65 \pm 0.07$ &        Point \\
1237661849863585864 & 200.125410 &  43.083937 & $1275.91 \pm 100.00$ & $12.29 \pm 0.01$ & Extended
\enddata 
\tablenotetext{a}{Radial velocities ($cz$) of stars. }
\tablenotetext{b}{Classification from SDSS database: Point means a point source and Extended indicates an extended source. }
\end{deluxetable}

\facilities{MMT (Hectospec), Sloan, Subaru (Hyper Suprime-Cam)}
\software{astropy \citep{astropy13, astropy18}} 

\bibliography{ms}{}

\begin{thebibliography}{}
\expandafter\ifx\csname natexlab\endcsname\relax\def\natexlab#1{#1}\fi
\providecommand{\url}[1]{\href{#1}{#1}}
\providecommand{\dodoi}[1]{doi:~\href{http://doi.org/#1}{\nolinkurl{#1}}}
\providecommand{\doeprint}[1]{\href{http://ascl.net/#1}{\nolinkurl{http://ascl.net/#1}}}
\providecommand{\doarXiv}[1]{\href{https://arxiv.org/abs/#1}{\nolinkurl{https://arxiv.org/abs/#1}}}

\bibitem[{{Ahumada} {et~al.}(2020){Ahumada}, {Prieto}, {Almeida}, {Anders},
  {Anderson}, {Andrews}, {Anguiano}, {Arcodia}, {Armengaud}, {Aubert}, {Avila},
  {Avila-Reese}, {Badenes}, {Balland}, {Barger}, {Barrera-Ballesteros}, {Basu},
  {Bautista}, {Beaton}, {Beers}, {Benavides}, {Bender}, {Bernardi}, {Bershady},
  {Beutler}, {Bidin}, {Bird}, {Bizyaev}, {Blanc}, {Blanton}, {Boquien},
  {Borissova}, {Bovy}, {Brandt}, {Brinkmann}, {Brownstein}, {Bundy}, {Bureau},
  {Burgasser}, {Burtin}, {Cano-D{\'\i}az}, {Capasso}, {Cappellari}, {Carrera},
  {Chabanier}, {Chaplin}, {Chapman}, {Cherinka}, {Chiappini}, {Doohyun Choi},
  {Chojnowski}, {Chung}, {Clerc}, {Coffey}, {Comerford}, {Comparat}, {da
  Costa}, {Cousinou}, {Covey}, {Crane}, {Cunha}, {Ilha}, {Dai}, {Damsted},
  {Darling}, {Davidson}, {Davies}, {Dawson}, {De}, {de la Macorra}, {De Lee},
  {Queiroz}, {Deconto Machado}, {de la Torre}, {Dell'Agli}, {du Mas des
  Bourboux}, {Diamond-Stanic}, {Dillon}, {Donor}, {Drory}, {Duckworth},
  {Dwelly}, {Ebelke}, {Eftekharzadeh}, {Davis Eigenbrot}, {Elsworth},
  {Eracleous}, {Erfanianfar}, {Escoffier}, {Fan}, {Farr},
  {Fern{\'a}ndez-Trincado}, {Feuillet}, {Finoguenov}, {Fofie},
  {Fraser-McKelvie}, {Frinchaboy}, {Fromenteau}, {Fu}, {Galbany}, {Garcia},
  {Garc{\'\i}a-Hern{\'a}ndez}, {Oehmichen}, {Ge}, {Maia}, {Geisler}, {Gelfand},
  {Goddy}, {Gonzalez-Perez}, {Grabowski}, {Green}, {Grier}, {Guo}, {Guy},
  {Harding}, {Hasselquist}, {Hawken}, {Hayes}, {Hearty}, {Hekker}, {Hogg},
  {Holtzman}, {Horta}, {Hou}, {Hsieh}, {Huber}, {Hunt}, {Chitham}, {Imig},
  {Jaber}, {Angel}, {Johnson}, {Jones}, {J{\"o}nsson}, {Jullo}, {Kim},
  {Kinemuchi}, {Kirkpatrick}, {Kite}, {Klaene}, {Kneib}, {Kollmeier}, {Kong},
  {Kounkel}, {Krishnarao}, {Lacerna}, {Lan}, {Lane}, {Law}, {Le Goff}, {Leung},
  {Lewis}, {Li}, {Lian}, {Lin}, {Long}, {Longa-Pe{\~n}a}, {Lundgren}, {Lyke},
  {Ted Mackereth}, {MacLeod}, {Majewski}, {Manchado}, {Maraston}, {Martini},
  {Masseron}, {Masters}, {Mathur}, {McDermid}, {Merloni}, {Merrifield},
  {M{\'e}sz{\'a}ros}, {Miglio}, {Minniti}, {Minsley}, {Miyaji}, {Mohammad},
  {Mosser}, {Mueller}, {Muna}, {Mu{\~n}oz-Guti{\'e}rrez}, {Myers}, {Nadathur},
  {Nair}, {Nandra}, {do Nascimento}, {Nevin}, {Newman}, {Nidever}, {Nitschelm},
  {Noterdaeme}, {O'Connell}, {Olmstead}, {Oravetz}, {Oravetz}, {Osorio},
  {Pace}, {Padilla}, {Palanque-Delabrouille}, {Palicio}, {Pan}, {Pan},
  {Parker}, {Paviot}, {Peirani}, {Ram{\'r}ez}, {Penny}, {Percival},
  {Perez-Fournon}, {P{\'e}rez-R{\`a}fols}, {Petitjean}, {Pieri},
  {Pinsonneault}, {Poovelil}, {Povick}, {Prakash}, {Price-Whelan}, {Raddick},
  {Raichoor}, {Ray}, {Rembold}, {Rezaie}, {Riffel}, {Riffel}, {Rix}, {Robin},
  {Roman-Lopes}, {Rom{\'a}n-Z{\'u}{\~n}iga}, {Rose}, {Ross}, {Rossi},
  {Rowlands}, {Rubin}, {Salvato}, {S{\'a}nchez}, {S{\'a}nchez-Menguiano},
  {S{\'a}nchez-Gallego}, {Sayres}, {Schaefer}, {Schiavon}, {Schimoia},
  {Schlafly}, {Schlegel}, {Schneider}, {Schultheis}, {Schwope}, {Seo},
  {Serenelli}, {Shafieloo}, {Shamsi}, {Shao}, {Shen}, {Shetrone}, {Shirley},
  {Aguirre}, {Simon}, {Skrutskie}, {Slosar}, {Smethurst}, {Sobeck}, {Sodi},
  {Souto}, {Stark}, {Stassun}, {Steinmetz}, {Stello}, {Stermer},
  {Storchi-Bergmann}, {Streblyanska}, {Stringfellow}, {Stutz}, {Su{\'a}rez},
  {Sun}, {Taghizadeh-Popp}, {Talbot}, {Tayar}, {Thakar}, {Theriault}, {Thomas},
  {Thomas}, {Tinker}, {Tojeiro}, {Toledo}, {Tremonti}, {Troup}, {Tuttle},
  {Unda-Sanzana}, {Valentini}, {Vargas-Gonz{\'a}lez}, {Vargas-Maga{\~n}a},
  {V{\'a}zquez-Mata}, {Vivek}, {Wake}, {Wang}, {Weaver}, {Weijmans}, {Wild},
  {Wilson}, {Wilson}, {Wolthuis}, {Wood-Vasey}, {Yan}, {Yang}, {Y{\`e}che},
  {Zamora}, {Zarrouk}, {Zasowski}, {Zhang}, {Zhao}, {Zhao}, {Zheng}, {Zheng},
  {Zhu}, \& {Zou}}]{SDSSDR16}
{Ahumada}, R., {Prieto}, C.~A., {Almeida}, A., {et~al.} 2020, \apjs, 249, 3,
  \dodoi{10.3847/1538-4365/ab929e}

\bibitem[{{Aihara} {et~al.}(2022){Aihara}, {AlSayyad}, {Ando}, {Armstrong},
  {Bosch}, {Egami}, {Furusawa}, {Furusawa}, {Harasawa}, {Harikane}, {Hsieh},
  {Ikeda}, {Ito}, {Iwata}, {Kodama}, {Koike}, {Kokubo}, {Komiyama}, {Li},
  {Liang}, {Lin}, {Lupton}, {Lust}, {MacArthur}, @{Mawatari}, {Mineo},
  {Miyatake}, {Miyazaki}, {More}, {Morishima}, {Murayama}, {Nakajima},
  {Nakata}, {Nishizawa}, {Oguri}, {Okabe}, {Okura}, {Ono}, {Osato}, {Ouchi},
  {Pan}, {Plazas Malag{\'o}n}, {Price}, {Reed}, {Rykoff}, {Shibuya},
  {Simunovic}, {Strauss}, {Sugimori}, {Suto}, {Suzuki}, {Takada}, {Takagi},
  {Takata}, {Takita}, {Tanaka}, {Tang}, {Taranu}, {Terai}, {Toba}, {Turner},
  {Uchiyama}, {Vijarnwannaluk}, {Waters}, {Yamada}, {Yamamoto}, \&
  {Yamashita}}]{Aihara22}
{Aihara}, H., {AlSayyad}, Y., {Ando}, M., {et~al.} 2022, \pasj, 74, 247,
  \dodoi{10.1093/pasj/psab122}

\bibitem[{{Arnouts} {et~al.}(1999){Arnouts}, {Cristiani}, {Moscardini},
  {Matarrese}, {Lucchin}, {Fontana}, \& {Giallongo}}]{Arnouts99}
{Arnouts}, S., {Cristiani}, S., {Moscardini}, L., {et~al.} 1999, \mnras, 310,
  540, \dodoi{10.1046/j.1365-8711.1999.02978.x}

\bibitem[{{Astropy Collaboration} {et~al.}(2013){Astropy Collaboration},
  {Robitaille}, {Tollerud}, {Greenfield}, {Droettboom}, {Bray}, {Aldcroft},
  {Davis}, {Ginsburg}, {Price-Whelan}, {Kerzendorf}, {Conley}, {Crighton},
  {Barbary}, {Muna}, {Ferguson}, {Grollier}, {Parikh}, {Nair}, {Unther},
  {Deil}, {Woillez}, {Conseil}, {Kramer}, {Turner}, {Singer}, {Fox}, {Weaver},
  {Zabalza}, {Edwards}, {Azalee Bostroem}, {Burke}, {Casey}, {Crawford},
  {Dencheva}, {Ely}, {Jenness}, {Labrie}, {Lim}, {Pierfederici}, {Pontzen},
  {Ptak}, {Refsdal}, {Servillat}, \& {Streicher}}]{astropy13}
{Astropy Collaboration}, {Robitaille}, T.~P., {Tollerud}, E.~J., {et~al.} 2013,
  \aap, 558, A33, \dodoi{10.1051/0004-6361/201322068}

\bibitem[{{Astropy Collaboration} {et~al.}(2018){Astropy Collaboration},
  {Price-Whelan}, {Sip{\H{o}}cz}, {G{\"u}nther}, {Lim}, {Crawford}, {Conseil},
  {Shupe}, {Craig}, {Dencheva}, {Ginsburg}, {VanderPlas}, {Bradley},
  {P{\'e}rez-Su{\'a}rez}, {de Val-Borro}, {Aldcroft}, {Cruz}, {Robitaille},
  {Tollerud}, {Ardelean}, {Babej}, {Bach}, {Bachetti}, {Bakanov}, {Bamford},
  {Barentsen}, {Barmby}, {Baumbach}, {Berry}, {Biscani}, {Boquien}, {Bostroem},
  {Bouma}, {Brammer}, {Bray}, {Breytenbach}, {Buddelmeijer}, {Burke},
  {Calderone}, {Cano Rodr{\'\i}guez}, {Cara}, {Cardoso}, {Cheedella}, {Copin},
  {Corrales}, {Crichton}, {D'Avella}, {Deil}, {Depagne}, {Dietrich}, {Donath},
  {Droettboom}, {Earl}, {Erben}, {Fabbro}, {Ferreira}, {Finethy}, {Fox},
  {Garrison}, {Gibbons}, {Goldstein}, {Gommers}, {Greco}, {Greenfield},
  {Groener}, {Grollier}, {Hagen}, {Hirst}, {Homeier}, {Horton}, {Hosseinzadeh},
  {Hu}, {Hunkeler}, {Ivezi{\'c}}, {Jain}, {Jenness}, {Kanarek}, {Kendrew},
  {Kern}, {Kerzendorf}, {Khvalko}, {King}, {Kirkby}, {Kulkarni}, {Kumar},
  {Lee}, {Lenz}, {Littlefair}, {Ma}, {Macleod}, {Mastropietro}, {McCully},
  {Montagnac}, {Morris}, {Mueller}, {Mumford}, {Muna}, {Murphy}, {Nelson},
  {Nguyen}, {Ninan}, {N{\"o}the}, {Ogaz}, {Oh}, {Parejko}, {Parley}, {Pascual},
  {Patil}, {Patil}, {Plunkett}, {Prochaska}, {Rastogi}, {Reddy Janga},
  {Sabater}, {Sakurikar}, {Seifert}, {Sherbert}, {Sherwood-Taylor}, {Shih},
  {Sick}, {Silbiger}, {Singanamalla}, {Singer}, {Sladen}, {Sooley},
  {Sornarajah}, {Streicher}, {Teuben}, {Thomas}, {Tremblay}, {Turner},
  {Terr{\'o}n}, {van Kerkwijk}, {de la Vega}, {Watkins}, {Weaver}, {Whitmore},
  {Woillez}, {Zabalza}, \& {Astropy Contributors}}]{astropy18}
{Astropy Collaboration}, {Price-Whelan}, A.~M., {Sip{\H{o}}cz}, B.~M., {et~al.}
  2018, \aj, 156, 123, \dodoi{10.3847/1538-3881/aabc4f}

\bibitem[{{Balogh} {et~al.}(1999){Balogh}, {Morris}, {Yee}, {Carlberg}, \&
  {Ellingson}}]{Balogh99}
{Balogh}, M.~L., {Morris}, S.~L., {Yee}, H.~K.~C., {Carlberg}, R.~G., \&
  {Ellingson}, E. 1999, \apj, 527, 54, \dodoi{10.1086/308056}

\bibitem[{{Blanton} \& {Moustakas}(2009)}]{Blanton09}
{Blanton}, M.~R., \& {Moustakas}, J. 2009, \araa, 47, 159,
  \dodoi{10.1146/annurev-astro-082708-101734}

\bibitem[{{Blanton} \& {Roweis}(2007)}]{Blanton07}
{Blanton}, M.~R., \& {Roweis}, S. 2007, \aj, 133, 734, \dodoi{10.1086/510127}

\bibitem[{{Bosch} {et~al.}(2018){Bosch}, {Armstrong}, {Bickerton}, {Furusawa},
  {Ikeda}, {Koike}, {Lupton}, {Mineo}, {Price}, {Takata}, {Tanaka}, {Yasuda},
  {AlSayyad}, {Becker}, {Coulton}, {Coupon}, {Garmilla}, {Huang}, {Krughoff},
  {Lang}, {Leauthaud}, {Lim}, {Lust}, {MacArthur}, {Mandelbaum}, {Miyatake},
  {Miyazaki}, {Murata}, {More}, {Okura}, {Owen}, {Swinbank}, {Strauss},
  {Yamada}, \& {Yamanoi}}]{Bosch18}
{Bosch}, J., {Armstrong}, R., {Bickerton}, S., {et~al.} 2018, \pasj, 70, S5,
  \dodoi{10.1093/pasj/psx080}

\bibitem[{{Bruzual} \& {Charlot}(2003)}]{BC03}
{Bruzual}, G., \& {Charlot}, S. 2003, \mnras, 344, 1000,
  \dodoi{10.1046/j.1365-8711.2003.06897.x}

\bibitem[{{Calzetti} {et~al.}(2000){Calzetti}, {Armus}, {Bohlin}, {Kinney},
  {Koornneef}, \& {Storchi-Bergmann}}]{Calzetti00}
{Calzetti}, D., {Armus}, L., {Bohlin}, R.~C., {et~al.} 2000, \apj, 533, 682,
  \dodoi{10.1086/308692}

\bibitem[{{Chabrier}(2003)}]{Chabrier03}
{Chabrier}, G. 2003, \pasp, 115, 763, \dodoi{10.1086/376392}

\bibitem[{{Conroy} {et~al.}(2009){Conroy}, {Gunn}, \& {White}}]{Conroy09}
{Conroy}, C., {Gunn}, J.~E., \& {White}, M. 2009, \apj, 699, 486,
  \dodoi{10.1088/0004-637X/699/1/486}

\bibitem[{{Dalton} {et~al.}(2012){Dalton}, {Trager}, {Abrams}, {Carter},
  {Bonifacio}, {Aguerri}, {MacIntosh}, {Evans}, {Lewis}, {Navarro}, {Agocs},
  {Dee}, {Rousset}, {Tosh}, {Middleton}, {Pragt}, {Terrett}, {Brock}, {Benn},
  {Verheijen}, {Cano Infantes}, {Bevil}, {Steele}, {Mottram}, {Bates},
  {Gribbin}, {Rey}, {Rodriguez}, {Delgado}, {Guinouard}, {Walton}, {Irwin},
  {Jagourel}, {Stuik}, {Gerlofsma}, {Roelfsma}, {Skillen}, {Ridings},
  {Balcells}, {Daban}, {Gouvret}, {Venema}, \& {Girard}}]{Dalton12}
{Dalton}, G., {Trager}, S.~C., {Abrams}, D.~C., {et~al.} 2012, in Society of
  Photo-Optical Instrumentation Engineers (SPIE) Conference Series, Vol. 8446,
  Ground-based and Airborne Instrumentation for Astronomy IV, ed. I.~S.
  {McLean}, S.~K. {Ramsay}, \& H.~{Takami}, 84460P, \dodoi{10.1117/12.925950}

\bibitem[{{Damjanov} {et~al.}(2022{\natexlab{a}}){Damjanov}, {Sohn}, {Geller},
  {Utsumi}, \& {Dell'Antonio}}]{Damjanov22b}
{Damjanov}, I., {Sohn}, J., {Geller}, M.~J., {Utsumi}, Y., \& {Dell'Antonio},
  I. 2022{\natexlab{a}}, arXiv e-prints, arXiv:2210.01129.
\newblock \doarXiv{2210.01129}

\bibitem[{{Damjanov} {et~al.}(2022{\natexlab{b}}){Damjanov}, {Sohn}, {Utsumi},
  {Geller}, \& {Dell'Antonio}}]{Damjanov22a}
{Damjanov}, I., {Sohn}, J., {Utsumi}, Y., {Geller}, M.~J., \& {Dell'Antonio},
  I. 2022{\natexlab{b}}, \apj, 929, 61, \dodoi{10.3847/1538-4357/ac54bd}

\bibitem[{{Damjanov} {et~al.}(2018){Damjanov}, {Zahid}, {Geller}, {Fabricant},
  \& {Hwang}}]{Damjanov18}
{Damjanov}, I., {Zahid}, H.~J., {Geller}, M.~J., {Fabricant}, D.~G., \&
  {Hwang}, H.~S. 2018, \apjs, 234, 21, \dodoi{10.3847/1538-4365/aaa01c}

\bibitem[{{Davies} {et~al.}(2018){Davies}, {Robotham}, {Driver}, {Lagos},
  {Cortese}, {Mannering}, {Foster}, {Lidman}, {Hashemizadeh}, {Koushan},
  {O'Toole}, {Baldry}, {Bilicki}, {Bland-Hawthorn}, {Bremer}, {Brown},
  {Bryant}, {Catinella}, {Croom}, {Grootes}, {Holwerda}, {Jarvis}, {Maddox},
  {Meyer}, {Moffett}, {Phillipps}, {Taylor}, {Windhorst}, \& {Wolf}}]{Davies18}
{Davies}, L.~J.~M., {Robotham}, A.~S.~G., {Driver}, S.~P., {et~al.} 2018,
  \mnras, 480, 768, \dodoi{10.1093/mnras/sty1553}

\bibitem[{{Davis} {et~al.}(1982){Davis}, {Huchra}, {Latham}, \&
  {Tonry}}]{Davis82}
{Davis}, M., {Huchra}, J., {Latham}, D.~W., \& {Tonry}, J. 1982, \apj, 253,
  423, \dodoi{10.1086/159646}

\bibitem[{{De Boni} {et~al.}(2016){De Boni}, {Serra}, {Diaferio}, {Giocoli}, \&
  {Baldi}}]{deBoni16}
{De Boni}, C., {Serra}, A.~L., {Diaferio}, A., {Giocoli}, C., \& {Baldi}, M.
  2016, \apj, 818, 188, \dodoi{10.3847/0004-637X/818/2/188}

\bibitem[{{Fabricant} {et~al.}(2005){Fabricant}, {Fata}, {Roll}, {Hertz},
  {Caldwell}, {Gauron}, {Geary}, {McLeod}, {Szentgyorgyi}, {Zajac}, {Kurtz},
  {Barberis}, {Bergner}, {Brown}, {Conroy}, {Eng}, {Geller}, {Goddard},
  {Honsa}, {Mueller}, {Mink}, {Ordway}, {Tokarz}, {Woods}, {Wyatt}, {Epps}, \&
  {Dell'Antonio}}]{Fabricant05}
{Fabricant}, D., {Fata}, R., {Roll}, J., {et~al.} 2005, \pasp, 117, 1411,
  \dodoi{10.1086/497385}

\bibitem[{{Fakhouri} {et~al.}(2010){Fakhouri}, {Ma}, \&
  {Boylan-Kolchin}}]{Fakhouri10}
{Fakhouri}, O., {Ma}, C.-P., \& {Boylan-Kolchin}, M. 2010, \mnras, 406, 2267,
  \dodoi{10.1111/j.1365-2966.2010.16859.x}

\bibitem[{{Geller} \& {Huchra}(1989)}]{Geller89}
{Geller}, M.~J., \& {Huchra}, J.~P. 1989, Science, 246, 897,
  \dodoi{10.1126/science.246.4932.897}

\bibitem[{{Geller} {et~al.}(2016){Geller}, {Hwang}, {Dell'Antonio}, {Zahid},
  {Kurtz}, \& {Fabricant}}]{Geller16}
{Geller}, M.~J., {Hwang}, H.~S., {Dell'Antonio}, I.~P., {et~al.} 2016, \apjs,
  224, 11, \dodoi{10.3847/0067-0049/224/1/11}

\bibitem[{{Geller} {et~al.}(2014){Geller}, {Hwang}, {Fabricant}, {Kurtz},
  {Dell'Antonio}, \& {Zahid}}]{Geller14}
{Geller}, M.~J., {Hwang}, H.~S., {Fabricant}, D.~G., {et~al.} 2014, \apjs, 213,
  35, \dodoi{10.1088/0067-0049/213/2/35}

\bibitem[{{Geller} {et~al.}(2010){Geller}, {Kurtz}, {Dell'Antonio}, {Ramella},
  \& {Fabricant}}]{Geller10}
{Geller}, M.~J., {Kurtz}, M.~J., {Dell'Antonio}, I.~P., {Ramella}, M., \&
  {Fabricant}, D.~G. 2010, \apj, 709, 832, \dodoi{10.1088/0004-637X/709/2/832}

\bibitem[{{Greene} {et~al.}(2022){Greene}, {Bezanson}, {Ouchi}, {Silverman}, \&
  {the PFS Galaxy Evolution Working Group}}]{Greene22}
{Greene}, J., {Bezanson}, R., {Ouchi}, M., {Silverman}, J., \& {the PFS Galaxy
  Evolution Working Group}. 2022, arXiv e-prints, arXiv:2206.14908.
\newblock \doarXiv{2206.14908}

\bibitem[{{Hahn} {et~al.}(2022){Hahn}, {Wilson}, {Ruiz-Macias}, {Cole},
  {Weinberg}, {Moustakas}, {Kremin}, {Tinker}, {Smith}, {Wechsler}, {Ahlen},
  {Alam}, {Bailey}, {Brooks}, {Cooper}, {Davis}, {Dawson}, {Dey}, {Dey},
  {Eftekharzadeh}, {Eisenstein}, {Fanning}, {Forero-Romero}, {Frenk},
  {Gazta{\~n}aga}, {Gontcho}, {Guy}, {Honscheid}, {Ishak}, {Juneau}, {Kehoe},
  {Kisner}, {Lan}, {Landriau}, {Le Guillou}, {Levi}, {Magneville}, {Martini},
  {Meisner}, {Myers}, {Nie}, {Norberg}, {Palanque-Delabrouille}, {Percival},
  {Poppett}, {Prada}, {Raichoor}, {Ross}, {Safonova}, {Saulder}, {Schlafly},
  {Schlegel}, {Sierra-Porta}, {Tarle}, {Weaver}, {Y{\`e}che}, {Zarrouk},
  {Zhou}, {Zhou}, \& {Zou}}]{Hahn22}
{Hahn}, C., {Wilson}, M.~J., {Ruiz-Macias}, O., {et~al.} 2022, arXiv e-prints,
  arXiv:2208.08512.
\newblock \doarXiv{2208.08512}

\bibitem[{{Hamadouche} {et~al.}(2022){Hamadouche}, {Carnall}, {McLure},
  {Dunlop}, {McLeod}, {Cullen}, {Begley}, {Bolzonella}, {Buitrago},
  {Castellano}, {Cucciati}, {Fontana}, {Gargiulo}, {Moresco}, {Pozzetti}, \&
  {Zamorani}}]{Hamadouche22}
{Hamadouche}, M.~L., {Carnall}, A.~C., {McLure}, R.~J., {et~al.} 2022, \mnras,
  512, 1262, \dodoi{10.1093/mnras/stac535}

\bibitem[{{Hsieh} \& {Yee}(2014)}]{Hsieh14}
{Hsieh}, B.~C., \& {Yee}, H.~K.~C. 2014, \apj, 792, 102,
  \dodoi{10.1088/0004-637X/792/2/102}

\bibitem[{{Huang} {et~al.}(2018{\natexlab{a}}){Huang}, {Leauthaud}, {Greene},
  {Bundy}, {Lin}, {Tanaka}, {Miyazaki}, \& {Komiyama}}]{Huang18b}
{Huang}, S., {Leauthaud}, A., {Greene}, J.~E., {et~al.} 2018{\natexlab{a}},
  \mnras, 475, 3348, \dodoi{10.1093/mnras/stx3200}

\bibitem[{{Huang} {et~al.}(2018{\natexlab{b}}){Huang}, {Leauthaud}, {Murata},
  {Bosch}, {Price}, {Lupton}, {Mandelbaum}, {Lackner}, {Bickerton}, {Miyazaki},
  {Coupon}, \& {Tanaka}}]{Huang18a}
{Huang}, S., {Leauthaud}, A., {Murata}, R., {et~al.} 2018{\natexlab{b}}, \pasj,
  70, S6, \dodoi{10.1093/pasj/psx126}

\bibitem[{{Hwang} {et~al.}(2016){Hwang}, {Geller}, {Park}, {Fabricant},
  {Kurtz}, {Rines}, {Kim}, {Diaferio}, {Zahid}, {Berlind}, {Calkins}, {Tokarz},
  \& {Moran}}]{Hwang16}
{Hwang}, H.~S., {Geller}, M.~J., {Park}, C., {et~al.} 2016, \apj, 818, 173,
  \dodoi{10.3847/0004-637X/818/2/173}

\bibitem[{{Ilbert} {et~al.}(2006){Ilbert}, {Arnouts}, {McCracken},
  {Bolzonella}, {Bertin}, {Le F{\`e}vre}, {Mellier}, {Zamorani}, {Pell{\`o}},
  {Iovino}, {Tresse}, {Le Brun}, {Bottini}, {Garilli}, {Maccagni}, {Picat},
  {Scaramella}, {Scodeggio}, {Vettolani}, {Zanichelli}, {Adami}, {Bardelli},
  {Cappi}, {Charlot}, {Ciliegi}, {Contini}, {Cucciati}, {Foucaud}, {Franzetti},
  {Gavignaud}, {Guzzo}, {Marano}, {Marinoni}, {Mazure}, {Meneux}, {Merighi},
  {Paltani}, {Pollo}, {Pozzetti}, {Radovich}, {Zucca}, {Bondi}, {Bongiorno},
  {Busarello}, {de La Torre}, {Gregorini}, {Lamareille}, {Mathez}, {Merluzzi},
  {Ripepi}, {Rizzo}, \& {Vergani}}]{Ilbert06}
{Ilbert}, O., {Arnouts}, S., {McCracken}, H.~J., {et~al.} 2006, \aap, 457, 841,
  \dodoi{10.1051/0004-6361:20065138}

\bibitem[{{Jaelani} {et~al.}(2020){Jaelani}, {More}, {Oguri}, {Sonnenfeld},
  {Suyu}, {Rusu}, {Wong}, {Chan}, {Kayo}, {Lee}, {Chao}, {Coupon}, {Inoue}, \&
  {Futamase}}]{Jaelani20}
{Jaelani}, A.~T., {More}, A., {Oguri}, M., {et~al.} 2020, \mnras, 495, 1291,
  \dodoi{10.1093/mnras/staa1062}

\bibitem[{{Kauffmann} {et~al.}(2004){Kauffmann}, {White}, {Heckman},
  {M{\'e}nard}, {Brinchmann}, {Charlot}, {Tremonti}, \&
  {Brinkmann}}]{Kauffmann04}
{Kauffmann}, G., {White}, S. D.~M., {Heckman}, T.~M., {et~al.} 2004, \mnras,
  353, 713, \dodoi{10.1111/j.1365-2966.2004.08117.x}

\bibitem[{{Kauffmann} {et~al.}(2003){Kauffmann}, {Heckman}, {White}, {Charlot},
  {Tremonti}, {Brinchmann}, {Bruzual}, {Peng}, {Seibert}, {Bernardi},
  {Blanton}, {Brinkmann}, {Castander}, {Cs{\'a}bai}, {Fukugita}, {Ivezic},
  {Munn}, {Nichol}, {Padmanabhan}, {Thakar}, {Weinberg}, \&
  {York}}]{Kauffmann03}
{Kauffmann}, G., {Heckman}, T.~M., {White}, S. D.~M., {et~al.} 2003, \mnras,
  341, 33, \dodoi{10.1046/j.1365-8711.2003.06291.x}

\bibitem[{{Kochanek} {et~al.}(2012){Kochanek}, {Eisenstein}, {Cool},
  {Caldwell}, {Assef}, {Jannuzi}, {Jones}, {Murray}, {Forman}, {Dey}, {Brown},
  {Eisenhardt}, {Gonzalez}, {Green}, \& {Stern}}]{Kochanek12}
{Kochanek}, C.~S., {Eisenstein}, D.~J., {Cool}, R.~J., {et~al.} 2012, \apjs,
  200, 8, \dodoi{10.1088/0067-0049/200/1/8}

\bibitem[{{Kurtz} {et~al.}(2012){Kurtz}, {Geller}, {Utsumi}, {Miyazaki},
  {Dell'Antonio}, \& {Fabricant}}]{Kurtz12}
{Kurtz}, M.~J., {Geller}, M.~J., {Utsumi}, Y., {et~al.} 2012, \apj, 750, 168,
  \dodoi{10.1088/0004-637X/750/2/168}

\bibitem[{{Kurtz} \& {Mink}(1998)}]{Kurtz98}
{Kurtz}, M.~J., \& {Mink}, D.~J. 1998, \pasp, 110, 934, \dodoi{10.1086/316207}

\bibitem[{{Lazo} {et~al.}(2018){Lazo}, {Zahid}, {Sohn}, \& {Geller}}]{Lazo18}
{Lazo}, B., {Zahid}, H.~J., {Sohn}, J., \& {Geller}, M.~J. 2018, Research Notes
  of the American Astronomical Society, 2, 234,
  \dodoi{10.3847/2515-5172/aaf8b1}

\bibitem[{{Liske} {et~al.}(2015){Liske}, {Baldry}, {Driver}, {Tuffs},
  {Alpaslan}, {Andrae}, {Brough}, {Cluver}, {Grootes}, {Gunawardhana},
  {Kelvin}, {Loveday}, {Robotham}, {Taylor}, {Bamford}, {Bland-Hawthorn},
  {Brown}, {Drinkwater}, {Hopkins}, {Meyer}, {Norberg}, {Peacock}, {Agius},
  {Andrews}, {Bauer}, {Ching}, {Colless}, {Conselice}, {Croom}, {Davies}, {De
  Propris}, {Dunne}, {Eardley}, {Ellis}, {Foster}, {Frenk}, {H{\"a}u{\ss}ler},
  {Holwerda}, {Howlett}, {Ibarra}, {Jarvis}, {Jones}, {Kafle}, {Lacey},
  {Lange}, {Lara-L{\'o}pez}, {L{\'o}pez-S{\'a}nchez}, {Maddox}, {Madore},
  {McNaught-Roberts}, {Moffett}, {Nichol}, {Owers}, {Palamara}, {Penny},
  {Phillipps}, {Pimbblet}, {Popescu}, {Prescott}, {Proctor}, {Sadler},
  {Sansom}, {Seibert}, {Sharp}, {Sutherland}, {V{\'a}zquez-Mata}, {van Kampen},
  {Wilkins}, {Williams}, \& {Wright}}]{Liske15}
{Liske}, J., {Baldry}, I.~K., {Driver}, S.~P., {et~al.} 2015, \mnras, 452,
  2087, \dodoi{10.1093/mnras/stv1436}

\bibitem[{{Maiolino} {et~al.}(2020){Maiolino}, {Cirasuolo}, {Afonso}, {Bauer},
  {Bowler}, {Cucciati}, {Daddi}, {De Lucia}, {Evans}, {Flores}, {Gargiulo},
  {Garilli}, {Jablonka}, {Jarvis}, {Kneib}, {Lilly}, {Looser}, {Magliocchetti},
  {Man}, {Mannucci}, {Maurogordato}, {McLure}, {Norberg}, {Oesch}, {Oliva},
  {Paltani}, {Pappalardo}, {Peng}, {Pentericci}, {Pozzetti}, {Renzini},
  {Rodrigues}, {Royer}, {Serjeant}, {Vanzi}, {Wild}, \&
  {Zamorani}}]{Maiolino20}
{Maiolino}, R., {Cirasuolo}, M., {Afonso}, J., {et~al.} 2020, The Messenger,
  180, 24, \dodoi{10.18727/0722-6691/5197}

\bibitem[{{Miyazaki} {et~al.}(2012){Miyazaki}, {Komiyama}, {Nakaya}, {Kamata},
  {Doi}, {Hamana}, {Karoji}, {Furusawa}, {Kawanomoto}, {Morokuma}, {Ishizuka},
  {Nariai}, {Tanaka}, {Uraguchi}, {Utsumi}, {Obuchi}, {Okura}, {Oguri},
  {Takata}, {Tomono}, {Kurakami}, {Namikawa}, {Usuda}, {Yamanoi}, {Terai},
  {Uekiyo}, {Yamada}, {Koike}, {Aihara}, {Fujimori}, {Mineo}, {Miyatake},
  {Yasuda}, {Nishizawa}, {Saito}, {Tanaka}, {Uchida}, {Katayama}, {Wang},
  {Chen}, {Lupton}, {Loomis}, {Bickerton}, {Price}, {Gunn}, {Suzuki},
  {Miyazaki}, {Muramatsu}, {Yamamoto}, {Endo}, {Ezaki}, {Itoh}, {Miwa},
  {Yokota}, {Matsuda}, {Ebinuma}, \& {Takeshi}}]{Miyazaki12}
{Miyazaki}, S., {Komiyama}, Y., {Nakaya}, H., {et~al.} 2012, in Society of
  Photo-Optical Instrumentation Engineers (SPIE) Conference Series, Vol. 8446,
  Ground-based and Airborne Instrumentation for Astronomy IV, ed. I.~S.
  {McLean}, S.~K. {Ramsay}, \& H.~{Takami}, 84460Z, \dodoi{10.1117/12.926844}

\bibitem[{{Petrosian}(1976)}]{Petrosian76}
{Petrosian}, V. 1976, \apjl, 210, L53, \dodoi{10.1086/182301}

\bibitem[{{Pizzardo} {et~al.}(2022){Pizzardo}, {Sohn}, {Geller}, {Diaferio}, \&
  {Rines}}]{Pizzardo22}
{Pizzardo}, M., {Sohn}, J., {Geller}, M.~J., {Diaferio}, A., \& {Rines}, K.
  2022, \apj, 927, 26, \dodoi{10.3847/1538-4357/ac5029}

\bibitem[{{Richard} {et~al.}(2019){Richard}, {Kneib}, {Blake}, {Raichoor},
  {Comparat}, {Shanks}, {Sorce}, {Sahl{\'e}n}, {Howlett}, {Tempel}, {McMahon},
  {Bilicki}, {Roukema}, {Loveday}, {Pryer}, {Buchert}, {Zhao}, \& {CRS
  Team}}]{Richard19}
{Richard}, J., {Kneib}, J.~P., {Blake}, C., {et~al.} 2019, The Messenger, 175,
  50, \dodoi{10.18727/0722-6691/5127}

\bibitem[{{Rines} {et~al.}(2016){Rines}, {Geller}, {Diaferio}, \&
  {Hwang}}]{Rines16}
{Rines}, K.~J., {Geller}, M.~J., {Diaferio}, A., \& {Hwang}, H.~S. 2016, \apj,
  819, 63, \dodoi{10.3847/0004-637X/819/1/63}

\bibitem[{{Robotham} {et~al.}(2010){Robotham}, {Driver}, {Norberg}, {Baldry},
  {Bamford}, {Hopkins}, {Liske}, {Loveday}, {Peacock}, {Cameron}, {Croom},
  {Doyle}, {Frenk}, {Hill}, {Jones}, {van Kampen}, {Kelvin}, {Kuijken},
  {Nichol}, {Parkinson}, {Popescu}, {Prescott}, {Sharp}, {Sutherland},
  {Thomas}, \& {Tuffs}}]{Robotham10}
{Robotham}, A., {Driver}, S.~P., {Norberg}, P., {et~al.} 2010, \pasa, 27, 76,
  \dodoi{10.1071/AS09053}

\bibitem[{{Ruiz-Macias} {et~al.}(2021){Ruiz-Macias}, {Zarrouk}, {Cole},
  {Baugh}, {Norberg}, {Lucey}, {Dey}, {Eisenstein}, {Doel}, {Gazta{\~n}aga},
  {Hahn}, {Kehoe}, {Kitanidis}, {Landriau}, {Lang}, {Moustakas}, {Myers},
  {Prada}, {Schubnell}, {Weinberg}, \& {Wilson}}]{RuizMacias21}
{Ruiz-Macias}, O., {Zarrouk}, P., {Cole}, S., {et~al.} 2021, \mnras, 502, 4328,
  \dodoi{10.1093/mnras/stab292}

\bibitem[{{Rykoff} {et~al.}(2016){Rykoff}, {Rozo}, {Hollowood},
  {Bermeo-Hernandez}, {Jeltema}, {Mayers}, {Romer}, {Rooney}, {Saro}, {Vergara
  Cervantes}, {Wechsler}, {Wilcox}, {Abbott}, {Abdalla}, {Allam}, {Annis},
  {Benoit-L{\'e}vy}, {Bernstein}, {Bertin}, {Brooks}, {Burke}, {Capozzi},
  {Carnero Rosell}, {Carrasco Kind}, {Castander}, {Childress}, {Collins},
  {Cunha}, {D'Andrea}, {da Costa}, {Davis}, {Desai}, {Diehl}, {Dietrich},
  {Doel}, {Evrard}, {Finley}, {Flaugher}, {Fosalba}, {Frieman}, {Glazebrook},
  {Goldstein}, {Gruen}, {Gruendl}, {Gutierrez}, {Hilton}, {Honscheid}, {Hoyle},
  {James}, {Kay}, {Kuehn}, {Kuropatkin}, {Lahav}, {Lewis}, {Lidman}, {Lima},
  {Maia}, {Mann}, {Marshall}, {Martini}, {Melchior}, {Miller}, {Miquel},
  {Mohr}, {Nichol}, {Nord}, {Ogando}, {Plazas}, {Reil}, {Sahl{\'e}n},
  {Sanchez}, {Santiago}, {Scarpine}, {Schubnell}, {Sevilla-Noarbe}, {Smith},
  {Soares-Santos}, {Sobreira}, {Stott}, {Suchyta}, {Swanson}, {Tarle},
  {Thomas}, {Tucker}, {Uddin}, {Viana}, {Vikram}, {Walker}, {Zhang}, \& {DES
  Collaboration}}]{Rykoff16}
{Rykoff}, E.~S., {Rozo}, E., {Hollowood}, D., {et~al.} 2016, \apjs, 224, 1,
  \dodoi{10.3847/0067-0049/224/1/1}

\bibitem[{{Shectman} {et~al.}(1996){Shectman}, {Landy}, {Oemler}, {Tucker},
  {Lin}, {Kirshner}, \& {Schechter}}]{Shectman96}
{Shectman}, S.~A., {Landy}, S.~D., {Oemler}, A., {et~al.} 1996, \apj, 470, 172,
  \dodoi{10.1086/177858}

\bibitem[{{Sohn} {et~al.}(2018{\natexlab{a}}){Sohn}, {Chon}, {B{\"o}hringer},
  {Geller}, {Diaferio}, {Hwang}, {Utsumi}, \& {Rines}}]{Sohn18b}
{Sohn}, J., {Chon}, G., {B{\"o}hringer}, H., {et~al.} 2018{\natexlab{a}}, \apj,
  855, 100, \dodoi{10.3847/1538-4357/aaac7a}

\bibitem[{{Sohn} {et~al.}(2021{\natexlab{a}}){Sohn}, {Geller}, {Hwang},
  {Diaferio}, {Rines}, \& {Utsumi}}]{Sohn21b}
{Sohn}, J., {Geller}, M.~J., {Hwang}, H.~S., {et~al.} 2021{\natexlab{a}}, \apj,
  923, 143, \dodoi{10.3847/1538-4357/ac29c3}

\bibitem[{{Sohn} {et~al.}(2021{\natexlab{b}}){Sohn}, {Geller}, {Hwang},
  {Fabricant}, {Moran}, \& {Utsumi}}]{Sohn21a}
---. 2021{\natexlab{b}}, \apj, 909, 129, \dodoi{10.3847/1538-4357/abd9be}

\bibitem[{{Sohn} {et~al.}(2018{\natexlab{b}}){Sohn}, {Geller}, {Rines},
  {Hwang}, {Utsumi}, \& {Diaferio}}]{Sohn18a}
{Sohn}, J., {Geller}, M.~J., {Rines}, K.~J., {et~al.} 2018{\natexlab{b}}, \apj,
  856, 172, \dodoi{10.3847/1538-4357/aab20b}

\bibitem[{{Sohn} {et~al.}(2022){Sohn}, {Geller}, {Vogelsberger}, \&
  {Borrow}}]{Sohn22}
{Sohn}, J., {Geller}, M.~J., {Vogelsberger}, M., \& {Borrow}, J. 2022, \apj,
  938, 3, \dodoi{10.3847/1538-4357/ac8f23}

\bibitem[{{Sohn} {et~al.}(2017{\natexlab{a}}){Sohn}, {Geller}, {Zahid},
  {Fabricant}, {Diaferio}, \& {Rines}}]{Sohn17a}
{Sohn}, J., {Geller}, M.~J., {Zahid}, H.~J., {et~al.} 2017{\natexlab{a}},
  \apjs, 229, 20, \dodoi{10.3847/1538-4365/aa653e}

\bibitem[{{Sohn} {et~al.}(2017{\natexlab{b}}){Sohn}, {Zahid}, \&
  {Geller}}]{Sohn17b}
{Sohn}, J., {Zahid}, H.~J., \& {Geller}, M.~J. 2017{\natexlab{b}}, \apj, 845,
  73, \dodoi{10.3847/1538-4357/aa7de3}

\bibitem[{{Strauss} {et~al.}(2002){Strauss}, {Weinberg}, {Lupton}, {Narayanan},
  {Annis}, {Bernardi}, {Blanton}, {Burles}, {Connolly}, {Dalcanton}, {Doi},
  {Eisenstein}, {Frieman}, {Fukugita}, {Gunn}, {Ivezi{\'c}}, {Kent}, {Kim},
  {Knapp}, {Kron}, {Munn}, {Newberg}, {Nichol}, {Okamura}, {Quinn}, {Richmond},
  {Schlegel}, {Shimasaku}, {SubbaRao}, {Szalay}, {Vanden Berk}, {Vogeley},
  {Yanny}, {Yasuda}, {York}, \& {Zehavi}}]{Strauss02}
{Strauss}, M.~A., {Weinberg}, D.~H., {Lupton}, R.~H., {et~al.} 2002, \aj, 124,
  1810, \dodoi{10.1086/342343}

\bibitem[{{Takada} {et~al.}(2014){Takada}, {Ellis}, {Chiba}, {Greene},
  {Aihara}, {Arimoto}, {Bundy}, {Cohen}, {Dor{\'e}}, {Graves}, {Gunn},
  {Heckman}, {Hirata}, {Ho}, {Kneib}, {Le F{\`e}vre}, {Lin}, {More},
  {Murayama}, {Nagao}, {Ouchi}, {Seiffert}, {Silverman}, {Sodr{\'e}},
  {Spergel}, {Strauss}, {Sugai}, {Suto}, {Takami}, \& {Wyse}}]{Takada14}
{Takada}, M., {Ellis}, R.~S., {Chiba}, M., {et~al.} 2014, \pasj, 66, R1,
  \dodoi{10.1093/pasj/pst019}

\bibitem[{{Tanaka} {et~al.}(2018){Tanaka}, {Coupon}, {Hsieh}, {Mineo},
  {Nishizawa}, {Speagle}, {Furusawa}, {Miyazaki}, \& {Murayama}}]{Tanaka18}
{Tanaka}, M., {Coupon}, J., {Hsieh}, B.-C., {et~al.} 2018, \pasj, 70, S9,
  \dodoi{10.1093/pasj/psx077}

\bibitem[{{Taylor} {et~al.}(2018){Taylor}, {Cirasuolo}, {Afonso}, {Carollo},
  {Evans}, {Flores}, {Maiolino}, {Paltani}, {Vanzi}, {Abreu}, {Amans},
  {Atkinson}, {Barrett}, {Beard}, {B{\'e}chet}, {Black}, {Boettger},
  {Brierley}, {Buscher}, {Cabral}, {Cochrane}, {Coelho}, {Colling},
  {Conzelmann}, {Dalessio}, {Dauvin}, {Davidson}, {Drass}, {D{\"u}nner},
  {Fairley}, {Fasola}, {Ferruzzi}, {Fisher}, {Flores}, {Garilli}, {Gaudemard},
  {Gonzalez}, {Guinouard}, {Gutierrez}, {Hammersley}, {Haigron}, {Haniff},
  {Hayati}, {Ives}, {Iwert}, {Laporte}, {Lee}, {Li Causi}, {Luco}, {Macleod},
  {Mainieri}, {Maire}, {Melse}, {Nix}, {Oliva}, {Oliveira}, {Origlia}, {Parry},
  {Pedichini}, {Piazzesi}, {Rees}, {Reix}, {Rodrigues}, {Rojas}, {Rota},
  {Royer}, {Santos}, {Schnell}, {Shen}, {Sordet}, {Strachan}, {Sun}, {Tait},
  {Torres}, {Tozzi}, {Tulloch}, {Navarro}, {Von Dran}, {Waring}, {Watson},
  {Woodward}, \& {Yang}}]{Taylor18}
{Taylor}, W., {Cirasuolo}, M., {Afonso}, J., {et~al.} 2018, in Society of
  Photo-Optical Instrumentation Engineers (SPIE) Conference Series, Vol. 10702,
  Ground-based and Airborne Instrumentation for Astronomy VII, ed. C.~J.
  {Evans}, L.~{Simard}, \& H.~{Takami}, 107021G, \dodoi{10.1117/12.2313403}

\bibitem[{{Utsumi} {et~al.}(2016){Utsumi}, {Geller}, {Dell'Antonio}, {Kamata},
  {Kawanomoto}, {Koike}, {Komiyama}, {Koshida}, {Mineo}, {Miyazaki}, {Sakurai},
  {Tait}, {Terai}, {Tomono}, {Usuda}, {Yamada}, \& {Zahid}}]{Utsumi16}
{Utsumi}, Y., {Geller}, M.~J., {Dell'Antonio}, I.~P., {et~al.} 2016, \apj, 833,
  156, \dodoi{10.3847/1538-4357/833/2/156}

\bibitem[{{Vergani} {et~al.}(2008){Vergani}, {Scodeggio}, {Pozzetti}, {Iovino},
  {Franzetti}, {Garilli}, {Zamorani}, {Maccagni}, {Lamareille}, {Le F{\`e}vre},
  {Charlot}, {Contini}, {Guzzo}, {Bottini}, {Le Brun}, {Picat}, {Scaramella},
  {Tresse}, {Vettolani}, {Zanichelli}, {Adami}, {Arnouts}, {Bardelli},
  {Bolzonella}, {Cappi}, {Ciliegi}, {Foucaud}, {Gavignaud}, {Ilbert},
  {McCracken}, {Marano}, {Marinoni}, {Mazure}, {Meneux}, {Merighi}, {Paltani},
  {Pell{\`o}}, {Pollo}, {Radovich}, {Zucca}, {Bondi}, {Bongiorno},
  {Brinchmann}, {Cucciati}, {de la Torre}, {Gregorini}, {Perez-Montero},
  {Mellier}, {Merluzzi}, \& {Temporin}}]{Vergani08}
{Vergani}, D., {Scodeggio}, M., {Pozzetti}, L., {et~al.} 2008, \aap, 487, 89,
  \dodoi{10.1051/0004-6361:20077910}

\bibitem[{{Wittman} {et~al.}(2006){Wittman}, {Dell'Antonio}, {Hughes},
  {Margoniner}, {Tyson}, {Cohen}, \& {Norman}}]{Wittman06}
{Wittman}, D., {Dell'Antonio}, I.~P., {Hughes}, J.~P., {et~al.} 2006, \apj,
  643, 128, \dodoi{10.1086/502621}

\bibitem[{{Woods} {et~al.}(2010){Woods}, {Geller}, {Kurtz}, {Westra},
  {Fabricant}, \& {Dell'Antonio}}]{Woods10}
{Woods}, D.~F., {Geller}, M.~J., {Kurtz}, M.~J., {et~al.} 2010, \aj, 139, 1857,
  \dodoi{10.1088/0004-6256/139/5/1857}

\bibitem[{{Zahid} {et~al.}(2019){Zahid}, {Geller}, {Damjanov}, \&
  {Sohn}}]{Zahid19}
{Zahid}, H.~J., {Geller}, M.~J., {Damjanov}, I., \& {Sohn}, J. 2019, \apj, 878,
  158, \dodoi{10.3847/1538-4357/ab21b9}

\bibitem[{{Zahid} {et~al.}(2016){Zahid}, {Geller}, {Fabricant}, \&
  {Hwang}}]{Zahid16}
{Zahid}, H.~J., {Geller}, M.~J., {Fabricant}, D.~G., \& {Hwang}, H.~S. 2016,
  \apj, 832, 203, \dodoi{10.3847/0004-637X/832/2/203}

\bibitem[{{Zhou} {et~al.}(2022){Zhou}, {Dey}, {Newman}, {Eisenstein}, {Dawson},
  {Bailey}, {Berti}, {Guy}, {Lan}, {Zou}, {Aguilar}, {Ahlen}, {Alam}, {Brooks},
  {de la Macorra}, {Dey}, {Dhungana}, {Fanning}, {Font-Ribera}, {Gontcho},
  {Honscheid}, {Ishak}, {Kisner}, {Kov{\'a}cs}, {Kremin}, {Landriau}, {Levi},
  {Magneville}, {Martini}, {Meisner}, {Miquel}, {Moustakas}, {Myers}, {Nie},
  {Palanque-Delabrouille}, {Percival}, {Poppett}, {Prada}, {Raichoor}, {Ross},
  {Schlafly}, {Schlegel}, {Schubnell}, {Tarl{\'e}}, {Weaver}, {Wechsler},
  {Y{\`e}che}, \& {Zhou}}]{Zhou22}
{Zhou}, R., {Dey}, B., {Newman}, J.~A., {et~al.} 2022, arXiv e-prints,
  arXiv:2208.08515.
\newblock \doarXiv{2208.08515}

\end{thebibliography}
\bibliographystyle{aasjournal}

\end{document}